\documentstyle{ioplppt}

\begin{document}
\jl{6}

%\baselineskip=24pt

%\draft

\title{Perturbations of spacetime: gauge transformations and
gauge invariance at second order and beyond}[Perturbations of spacetime]

\author{Marco Bruni\dag\ddag, Sabino Matarrese\S, Silvia
Mollerach\ddag\P\ and Sebastiano Sonego\ddag}
\address{\dag\ ICTP -- International Center for Theoretical Physics,
P.~O.~Box 586, 34014 Trieste, Italy}
\address{\ddag\ SISSA -- International School for Advanced Studies,
Via Beirut 2--4, 34014 Trieste, Italy}
\address{\S\ Dipartimento di Fisica ``G. Galilei'', Universit\`a di
Padova, via Marzolo 8, 35131 Padova, Italy} 
\address{\P\ Departamento de Astronomia y Astrof\`\ii sica,
 Universidad de Valencia, 46100 Burjassot, Valencia, Spain}

%\date{}

\begin{abstract}

We consider in detail the problem of gauge dependence that exists in
relativistic perturbation theory, going beyond the linear
approximation and treating second and higher order perturbations.
We first derive some mathematical results concerning the Taylor
expansion of tensor fields under the action of one-parameter families
(not necessarily groups) of diffeomorphisms. Second, we define gauge
invariance to an arbitrary order $n$. Finally, we give a generating
formula for the gauge transformation to an arbitrary order and
explicit rules to second and third order.  This formalism can be used
in any field of applied general relativity, such as cosmological and
black hole perturbations, as well as in other spacetime theories. As a
specific example, we consider here second order perturbations in
cosmology, assuming a flat Robertson--Walker background, giving
explicit second order transformations between the synchronous and the
Poisson (generalized longitudinal) gauges.

\end{abstract}
\maketitle

%%%%%%%%%%%%%%%%%%%%%%%%%%%%%%%%%%%%%%%%%%%%%%%%%%%%%%%%%%%%%%%%%%%%%%
%%%%%%%%%%%%%%%%%%%%%%%%%%%%%%%%%%%%%%%%%%%%%%%%%%%%%%%%%%%%%%%%%%%%%%
%%%%%%%%%%%%%%%%%%%%%%%%%%%%%%%%%%%%%%%%%%%%%%%%%%%%%%%%%%%%%%%%%%%%%%

\section{Introduction}

\setcounter{equation}{0}

The perturbative approach is a fundamental tool of investigation in
general relativity, where exact solutions are most often too idealized
to properly represent the realm of natural phenomena.  Unfortunately,
it has long been known that the invariance of general relativity under
diffeomorphisms (two solutions of the Einstein equation are physically
equivalent if they are diffeomorphic to each other) makes the very
definition of perturbations gauge dependent
 \cite{bi:sachs,bi:SW,bi:JB,bi:KS,bi:EB,bi:stewart,bi:sb,bi:MFB,bi:RD}.
A gauge choice is an identification between points of the perturbed
and the background spacetimes, and generic perturbations are not
invariant under a gauge transformation.  This ``gauge problem'' has
been widely treated in linear theory, but what if one wants to
consider higher order perturbations?  In particular, how do the latter
change under a gauge transformation?

Second order treatments have been recently proposed, both in cosmology
\cite{bi:MPS,bi:russ} and compact object theory \cite{bi:gleiseretal},
as a way of obtaining more accurate results to be compared with
present and future observations. For example, in view of the increased
sensitivity expected from the next generation of detectors, precise
computations of microwave background anisotropies or gravitational
wave production may require going beyond the linear regime.  In
addition, second order perturbations provide a reliable measure of the
accuracy of the linearized theory (see, e.g., \cite{bi:gleiseretal}).
In cosmology, a second (or higher) order treatment might prove
necessary when dealing with scales much smaller than the cosmological
horizon, where the linear approximation can either be not accurate
enough or just miss some physical effects.  Finally, general
relativity is an intrinsically non-linear theory, thus it is in
principle interesting to look at higher order perturbations as a tool
for exploring non-linear features.  Unfortunately, already the second
order calculations are almost invariably a computational {\it tour de
force\/}, and a gauge-invariant treatment is not at hand.  As a matter
of fact, the computationally more convenient gauge does not
necessarily coincide with the most interesting one, and often
different authors work in different gauges, but at present a general
formalism to deal with second and higher order gauge transformations
is not available in the literature,  although some partial results
can be found in \cite{taub,bi:GL,bi:FW} and in \cite{bi:schutz}
(and references therein).

The aim of this paper is to fill this gap. Thus, while we shall
consider the problem of gauge transformations from a general
geometrical perspective, from the practical point of view our main
goal is to derive the effect on a tensor field $T$ of a second order
gauge transformation. To this end, we shall show that the latter is
necessarily represented in coordinates by
\begin{equation}
\label{eq:t}
\tilde{x}^\mu =x^\mu +\lambda\,\xi_{(1)}^\mu +
\frac{\lambda^2}{2}\left(\xi_{(1),\nu}^\mu \xi_{(1)}^\nu +
\xi_{(2)}^\mu\right)+ \cdots \,,
\end{equation}
where $\xi_{(1)}$ and $\xi_{(2)}$ are two independent vector
fields, and that, under (\ref{eq:t}),
 the first and second order perturbations of
$T$ transform as
\begin{eqnarray}
\label{eq:t1}
\delta \tilde{T} &=&\delta T +\pounds_{\xi_{(1)}} T_{0}\, ,\\
\label{eq:t2} 
\delta^2 \tilde{T}& =&\delta^2 T +2\pounds_{\xi_{(1)}} \delta T
+\pounds^{2}_{\xi_{(1)}} T_{0} +\pounds_{\xi_{(2)}}
T_{0}\;.
\end{eqnarray}
 Equation (\ref{eq:t1}) is the usual first order result in terms of
the Lie derivative of the background tensor field $T_0$ along the
vector field $\xi_{(1)}$, and the second order perturbation
$\delta^2T$ is defined by (\ref{DeltaT1}) and (\ref{deltakT}).
In fact, we derive a general formula, equation (\ref{lemma2applied}), from
which the gauge transformations to an arbitrary order $n$ can be deduced,
although we will give the explicit expression only for transformations
up to third order.  Furthermore, we show that a tensor field $T$ is
gauge-invariant to second order if and only if it is gauge-invariant
to first order, i.e.,
\begin{equation}
\pounds_{\xi} T_{0}=0\, 
\end{equation}
for an arbitrary vector field $\xi$, and, in addition,
\begin{equation}
\pounds_{\xi} \delta T =0\, .
\end{equation}
Actually, 
we generalize the above result giving  a condition
for gauge invariance to an arbitrary order $n$.

The plan of the paper is the following.  In the next section we give
some mathematical results concerning the Taylor expansion of tensor
fields under the action of one-parameter families (not necessarily
groups) of diffeomorphisms.  This material is necessary for the
applications to follow; however, the reader interested only in the
latter can skip all the proofs, as well as the last two paragraphs of
section \ref{secIIb};
 section \ref{secIId} is useful if one wants to translate our results in the 
 language of coordinates.
In section 3 we give
a general discussion of spacetime perturbations and gauge choices,
giving in particular a precise definition of the $k$-th order
perturbation of a general tensor field.  In section 4 we first define
total gauge invariance and gauge invariance to order $n$; then we give
the generating formula for a gauge transformation to an arbitrary
order, and explicit rules for second and third order
transformations.\footnote{A short account of the material covered in
sections 2--4 is presented in \cite{bi:roma}.}  In section 5 we
consider perturbations of a flat Robertson--Walker model, and apply
our results to the case of the transformations between two specific
gauge choices, i.e., the synchronous \cite{bi:li46} and the Poisson
\cite{bi:bertschinger} gauges. Section 6 contains a final discussion.

In general, we shall work on an $m$-dimensional Lorentzian manifold of
signature \hbox{$m-2$}.  The abstract mathematical notation is used
extensively, but sometimes we shall make reference to charts; in this
case, coordinate indices $\mu,\nu,\ldots$ take values from 0 to $m-1$.
Units are such that $c=1$.  Furthermore, for the sake of simplicity we
shall always suppose that manifolds, maps, tensor fields, etc., are as
smooth as necessary, or even analytic (see \cite{bi:sm} for an
extension to $C^r$ fields).

%%%%%%%%%%%%%%%%%%%%%%%%%%%%%%%%%%%%%%%%%%%%%%%%%%%%%%%%%%%%%%%%%%%%%%
%%%%%%%%%%%%%%%%%%%%%%%%%%%%%%%%%%%%%%%%%%%%%%%%%%%%%%%%%%%%%%%%%%%%%%
%%%%%%%%%%%%%%%%%%%%%%%%%%%%%%%%%%%%%%%%%%%%%%%%%%%%%%%%%%%%%%%%%%%%%%

\section{Taylor expansions of tensor fields}

\setcounter{equation}{0}

Before considering the specific problem of gauge transformations in
relativistic perturbation theory we need to establish some general
results concerning the Taylor expansion of tensor fields, to be used
later.  For functions on ${\rm I\!R}^m$, a Taylor expansion is
essentially a convenient way to express the value of the function at
some point in terms of its value, and the value of all its
derivatives, at another point.  Of course, this is impossible for a
tensor field $T$ on a manifold $\cal M$, simply because $T(p)$ and
$T(q)$ at different points $p$ and $q$ belong to different spaces, and
cannot thus be directly compared.  A Taylor expansion can therefore be
written only if one is given a mapping between tensors at different
points of $\cal M$.  In this section, we study the case in which such
a mapping arises from a one-parameter {\it family\/} of
diffeomorphisms of $\cal M$, starting from the simplest case of a flow
(i.e., a one-parameter {\it group\/} of diffeomorphisms) and then
proceeding to generalize it.  Let us first establish some notation.

Let $\varphi:{\cal M}\to{\cal N}$ be a diffeomorphism between two
manifolds $\cal M$ and $\cal N$.  For each $p\in{\cal M}$, $\varphi$
defines naturally the linear map $\left.\varphi_*\right|_p:T_p{\cal M}
\to T_{\varphi(p)}{\cal N}$ between the tangent spaces, called {\em
push-forward\/}, and the linear map
$\left.\varphi^*\right|_{\varphi(p)}:T^*_{\varphi(p)}{\cal N}\to
T^*_p{\cal M}$ between the cotangent spaces, called the {\em
pull-back\/}.\footnote{Here, we are following the most common
notation, although some authors (see, e.g., \cite{bi:waldbook})
denote the push-forward and the pull-back exactly in the opposite
way.}  Using $\varphi^{-1}$, we can define also a push-forward of
$T^*_p{\cal M}$ on $T^*_{\varphi(p)}{\cal N}$, and a pull-back of
$T_{\varphi(p)}{\cal N}$ on $T_p{\cal M}$, so that
$\left.\varphi_*\right|_p$ and $\left.\varphi^*\right|_{\varphi(p)}$
turn out to be well-defined for tensors of arbitrary type.  We can
also construct maps $\varphi_*$ and $\varphi^*$ for tensor fields,
simply requiring that, $\forall\,p\in{\cal M}$,
\begin{equation}
\left(\varphi_*T\right)(\varphi(p)):=
\left.\varphi_*\right|_p\left(T(p)\right)
\label{phi_*}
\end{equation}
and
\begin{equation}
\left(\varphi^*T\right)(p):=
\left.\varphi^*\right|_{\varphi(p)}\left(T(\varphi(p))\right)\;.
\label{phi^*}
\end{equation}
Hereafter, we drop the suffixes in $\left.\varphi_*\right|_p$ and
$\left.\varphi^*\right|_{\varphi(p)}$, since there is no real danger
of confusion.

%%%%%%%%%%%%%%%%%%%%%%%%%%%%%%%%%%%%%%%%%%%%%%%%%%%%%%%%%%%%%%%%%%%%%%
%%%%%%%%%%%%%%%%%%%%%%%%%%%%%%%%%%%%%%%%%%%%%%%%%%%%%%%%%%%%%%%%%%%%%%

\subsection{Flows}

Let ${\cal M}$ be a differentiable manifold, and let $\xi$ be a vector
field on ${\cal M}$, generating a flow $\phi:{\rm I\!R}\times{\cal
M}\to{\cal M}$, where $\phi(0,p)=p$, $\forall\,p\in{\cal
M}$.\footnote{In order not to burden the discussion unnecessarily, we
suppose that $\phi$ defines global transformations of $\cal M$
 \cite{bi:ycm}.}  For any given $\lambda\in{\rm I\!R}$, we shall write,
following the common usage, $\phi_\lambda(p):=\phi(\lambda,p)$,
$\forall\, p\in{\cal M}$.  Let $T$ be a tensor field on ${\cal M}$.
The map $\phi_{\lambda}^*$ defines a new field $\phi^*_{\lambda} T$ on
${\cal M}$, the pull-back of $T$, which is thus a function of
$\lambda$. \\

\noindent
{\bf Lemma 1:} The field $\phi^*_{\lambda} T$ 
admits the following expansion around $\lambda=0$:
\begin{equation} \phi^*_{\lambda} T=\sum^{+\infty}_{k=0}
\,\frac{\lambda^k}{k!}\,\pounds^k_\xi T\;.
\label{lemma1}
\end{equation}

\noindent
{\bf Proof:}  By analyticity we have
\begin{equation}
\phi^*_{\lambda} T = \sum^{+\infty}_{k=0}\, \frac{\lambda^k}{k!}\, 
\left.\frac{{\rm d}^k~}{{\rm d} \lambda^k}\right|_0  \phi^*_{\lambda}
T\;,
\end{equation}
where, here and in the following, 
\begin{equation}
\left.\frac{{\rm d}^k~}{{\rm d} \lambda^k}\right|_\tau (\cdots):=
\left[\frac{{\rm d}^k(\cdots)}{{\rm d} 
\lambda^k}\right]_{\lambda=\tau}\;,
\end{equation} 
and the first derivative is, by definition, just the Lie derivative of
$T$ with respect to $\xi$:
\begin{equation}
 \left. \frac{{\rm d}~}{{\rm d}\lambda}\right|_0  \phi^*_{\lambda}
T=\lim_{\lambda\rightarrow 0}\frac{1}{\lambda}\,
\left(\phi_{\lambda}^*T-T\right) 
=:\pounds_\xi T\;.
\end{equation}
In order to prove (\ref{lemma1}), 
it is then sufficient to show that, 
$\forall\,k$,
\begin{equation}
\left.\frac{{\rm d}^k~}{{\rm d} 
\lambda^k}\right|_0  \phi^*_{\lambda}
T=\pounds_\xi^kT\;.
\label{??}
\end{equation}
This can be established by induction over $k$.  Suppose that (\ref{??}) is 
true for some $k$.  Then
\begin{eqnarray}
\lefteqn{ \left. \frac{{\rm d}^{k+1}~}{{\rm d} 
\lambda^{k+1}}\right|_0 
\phi^*_{\lambda} T =}\nonumber\\ 
&& \lim_{\varepsilon\rightarrow 0}
\frac{1}{\varepsilon} \left( \left. \frac{{\rm d}^{k}~}{{\rm d}
\lambda^{k}}\right|_\varepsilon 
\phi^*_{\lambda} T -\left. 
\frac{{\rm d}^{k}~}{{\rm d}\lambda^{k}}\right|_0 
\phi^*_{\lambda} T\right)=\nonumber\\ 
&& \lim_{\varepsilon\rightarrow 0}
\frac{1}{\varepsilon} \left( \left. 
\frac{{\rm d}^{k}~}{{\rm d}\tau^{k}}\right|_0 
\phi^*_{\tau+\varepsilon} T -\left. \frac{{\rm d}^{k}~}{{\rm
d}\lambda^{k}}\right|_0 \phi^*_{\lambda} T\right)=\nonumber\\
&&\lim_{\varepsilon\rightarrow 0}
\frac{1}{\varepsilon} \left( \phi^*_{\varepsilon}\pounds_\xi^k 
T- \pounds_\xi^k T\right) = \pounds_\xi^{k+1}T\;,
\end{eqnarray}
where $\tau:=\lambda-\varepsilon$,
 and we have used the property that
$\phi_\lambda$ forms a one-parameter group:
$\phi_{\tau+\varepsilon}=\phi_\tau \circ \phi_\varepsilon$.
\hfill$\Box$\\

\noindent
It is worth noticing that (\ref{lemma1}) can also be
written  \cite{bi:schouten} in the symbolic form $\phi^*_\lambda
=\exp(\lambda\pounds_\xi)$. 

Equation (\ref{lemma1}) can be applied to the special case in which
the tensor $T$ is just one of the coordinate functions on $\cal M$,
$x^\mu$.  We have then, since
$\phi^*_{\lambda}x^\mu(p)=x^\mu(\phi_\lambda(p))$, the usual action of
an ``infinitesimal point transformation'' extended to second order in
$\lambda$:
\begin{equation} \tilde{x}^\mu=x^\mu+\lambda\,\xi^\mu+{\lambda^2\over
2}\,{\xi^\mu}_{,\nu}\xi^\nu+\cdots\;;
\label{lemma1coord}
\end{equation}
where we have denoted $x^\mu(p)$ simply by $x^\mu$, and
$x^\mu(\phi_\lambda(p))$ by $\tilde{x}^\mu$.

%%%%%%%%%%%%%%%%%%%%%%%%%%%%%%%%%%%%%%%%%%%%%%%%%%%%%%%%%%%%%%%%%%%%%%
%%%%%%%%%%%%%%%%%%%%%%%%%%%%%%%%%%%%%%%%%%%%%%%%%%%%%%%%%%%%%%%%%%%%%%

\subsection{Knight diffeomorphisms}
\label{secIIb}

\begin{figure}
\vspace{310pt}
\includegraphics{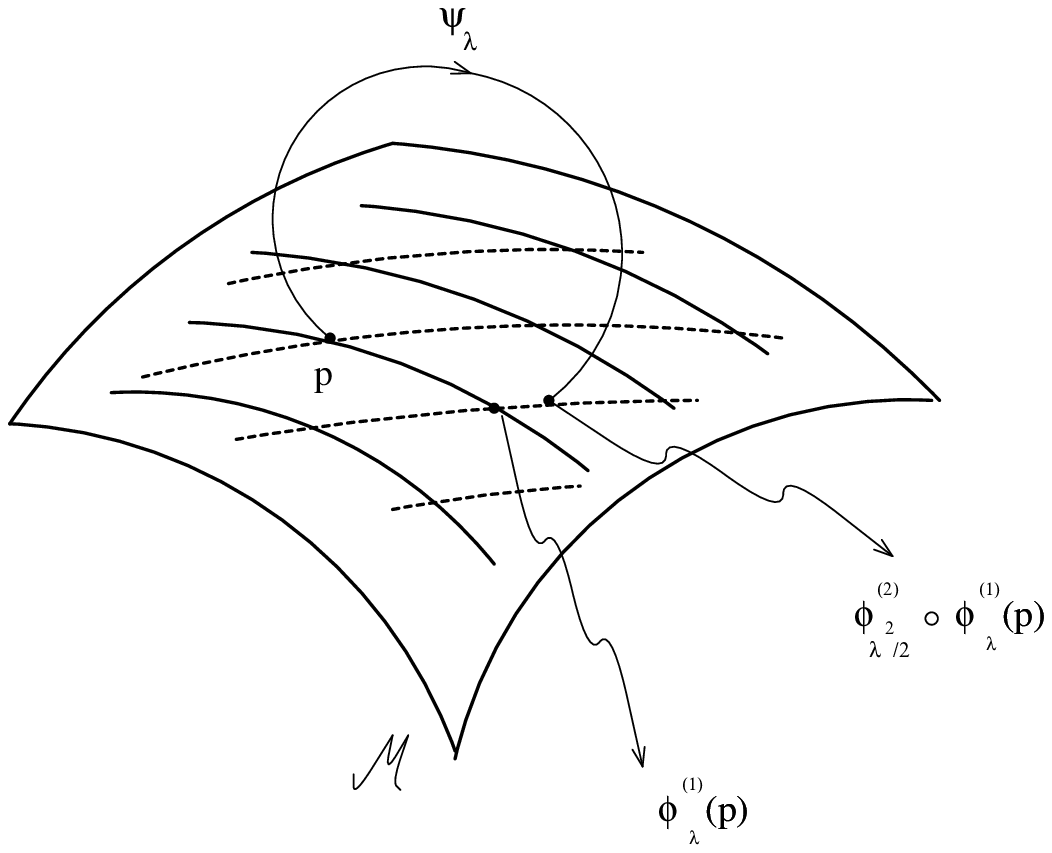}
\caption{The action of a knight diffeomorphism $\Psi_\lambda$
 generated by $\xi_{(1)}$
and $\xi_{(2)}$. Solid lines: integral curves of $\xi_{(1)}$. Dashed
lines: integral curves of $\xi_{(2)}$. The parameter lapse between
$p$ and $ \phi^{(1)}_\lambda(p)$ is $\lambda$, and that from $
\phi^{(1)}_\lambda(p)$ to $\phi^{(2)}_{\lambda^2/2}\circ
\phi^{(1)}_\lambda(p)$ is $\lambda^2/2$.}\label{fig:knight}
\end{figure}

Let us now suppose that there are {\em two\/} vector fields
$\xi_{(1)}$ and $\xi_{(2)}$ on ${\cal M}$.  Separately, they generate
the flows $\phi^{(1)}$ and $\phi^{(2)}$, respectively.  We can combine
$\phi^{(1)}$ and $\phi^{(2)}$ to define a new one-parameter family of
diffeomorphisms $\Psi:{\rm I\!R}\times{\cal M}\to{\cal M}$, whose
action is given by $\Psi_\lambda :=
\phi^{(2)}_{\lambda^2/2}\circ\phi^{(1)}_\lambda$. Thus, $\Psi_\lambda$
displaces a point of ${\cal M}$ a parameter interval $\lambda$ along
the integral curve of $\xi_{(1)}$, and then an interval $\lambda^2/2$
along the integral curve of $\xi_{(2)}$ (see figure \ref{fig:knight}).
For this reason, we shall call it, with a chess-inspired terminology,
a {\em knight diffeomorphism.\/}

This concept can be immediately generalized to the case in which $n$
vector fields $\xi_{(1)},\ldots,\xi_{(n)}$ are defined on ${\cal M}$,
corresponding to the flows $\phi^{(1)},\ldots, \phi^{(n)}$.  Then we
define a one-parameter family $\Psi:{\rm I\!R}\times{\cal M}\to{\cal
M}$ of knight diffeomorphisms of rank $n$ by
\begin{equation}\label{eq:knightn}
\Psi_\lambda:=\phi^{(n)}_{\lambda^n/n!}
\circ\cdots\circ\phi^{(2)}_{\lambda^2/2}
\circ\phi^{(1)}_\lambda\;,
\end{equation}
and the vector fields $\xi_{(1)},\ldots,\xi_{(n)}$ will be called the
{\em generators\/} of $\Psi$.

Of course, $\Psi_\sigma\circ\Psi_\lambda\neq\Psi_{\sigma+\lambda}$;
consequently, Lemma 1 cannot be applied if we want to expand in
$\lambda$ the pull-back $\Psi_{\lambda}^* T$ of a tensor field $T$
defined on ${\cal M}$.  However, the result can be easily
generalized.\\

\noindent
{\bf Lemma 2:} The pull-back 
$\Psi_{\lambda}^* T$ of a tensor field $T$ by a
one-parameter family of knight diffeomorphisms
$\Psi$ with generators $\xi_{(1)},\ldots,\xi_{(k)},
\ldots$ can be expanded
around $\lambda=0$ as follows: 
\begin{eqnarray}
\fl \Psi_{\lambda}^* T=\sum_{l_1=0}^{+\infty}
 \sum_{l_2=0}^{+\infty}\cdots
\sum_{l_k=0}^{+\infty}\cdots \,
 {\lambda^{l_1+2l_2+\cdots+kl_k+\cdots}\over
2^{l_2}\cdots (k!)^{l_k}\cdots l_1!l_2!\cdots
l_k!\cdots}\,\pounds^{l_1}_{\xi_{(1)}}\pounds^{l_2}_{\xi_{(2)}}
\cdots\pounds^{l_k}_{\xi_{(k)}}\cdots
T\;.\nonumber\\
\label{lemma2}
\end{eqnarray}

\noindent
{\bf Proof:}
\begin{eqnarray}
\fl \Psi_{\lambda}^* T&=\phi^{(1)*}_{\lambda} 
\phi^{(2)*}_{\lambda^2/2}\cdots
\phi^{(k)*}_{\lambda^k/k!}\cdots T=
\sum_{l_1=0}^{+\infty} {\lambda^{l_1}\over
l_1!}\,\pounds^{l_1}_{\xi_{(1)}}
\left(\phi^{(2)*}_{\lambda^2/2}\cdots
\phi^{(k)*}_{\lambda^k/k!}\cdots T\right)
\nonumber\\ 
\fl &= \sum_{l_1=0}^{+\infty} \sum_{l_2=0}^{+\infty}\cdots
\sum_{l_k=0}^{+\infty}\cdots \,
 {\lambda^{l_1+2l_2+\cdots+kl_k+\cdots}\over
2^{l_2}\cdots (k!)^{l_k}\cdots l_1!l_2!\cdots
l_k!\cdots}\,\pounds^{l_1}_{\xi_{(1)}}\pounds^{l_2}_{\xi_{(2)}}
\cdots\pounds^{l_k}_{\xi_{(k)}}\cdots
T\;,\nonumber\\
\fl &
\end{eqnarray}
where we have repeatedly used (\ref{lemma1}).\hfill$\Box$\\

The explicit form of (\ref{lemma2}) up to the third order in
$\lambda$ is
\begin{eqnarray}
\Psi_{\lambda}^* T=&T+\lambda \pounds_{\xi_{(1)}} T
+\frac{\lambda^2}{2}\left(\pounds^2_{\xi_{(1)}}
+\pounds_{\xi_{(2)}}\right)T\nonumber\\
&+{\lambda^3\over
3!}\,\left(\pounds^3_{\xi_{(1)}}
+3\,\pounds_{\xi_{(1)}}\pounds_{\xi_{(2)}}+
\pounds_{\xi_{(3)}}\right)T+\cdots\;.
\label{lemma2explic}
\end{eqnarray}

\begin{figure}
\vspace{240pt}
\includegraphics{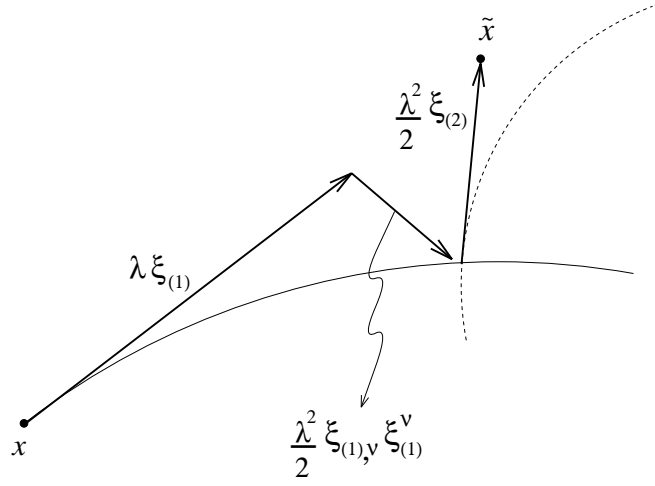}
\caption{The action of a knight diffeomorphism of rank two,
represented in a chart to second order.}\label{fig:chart}
\end{figure}

Equations (\ref{lemma2}) and (\ref{lemma2explic}) apply to a
one-parameter family of knight diffeomorphisms of arbitrarily high
rank, and can be specialized to the particular case of rank $n$ simply
by setting $\xi_{(k)}\equiv 0$, $\forall\, k>n$.  Their meaning is
particularly clear in a chart.  Denoting, as before, by $x^\mu$ the
coordinates of a point $p$, and by $\tilde{x}^\mu$ those of
$\Psi_\lambda(p)$, we have, to order $\lambda^2$,
\begin{equation} \tilde{x}^\mu=x^\mu
+\lambda\,\xi_{(1)}^\mu+{\lambda^2\over
2}\,\left({\xi_{(1)}^\mu}_{,\nu}\xi_{(1)}^\nu
+\xi_{(2)}^\mu\right)+\cdots\;.
\label{lemma2coord}
\end{equation}
Equation (\ref{lemma2coord}) is represented pictorially in 
figure \ref{fig:chart}. 

Since $\Psi_\sigma\circ\Psi_\lambda\neq\Psi_{\sigma+\lambda}$, and
$\Psi_\lambda^{-1}\neq\Psi_{-\lambda}$, one may reasonably doubt that
$\Psi$ forms a group, except under very special conditions.  This is
confirmed by the following\\

\noindent
{\bf Theorem 1:} The only cases in which $\Psi$ forms a group are
those for which $\xi_{(k)}=\alpha_k\xi_{(1)}$, $\forall\,k\geq 2$,
with $\alpha_k$ arbitrary numerical coefficients.  Then, under the
reparametrization $\bar{\lambda}:=f(\lambda)$, with
$f(\lambda):=\lambda+\sum_{k=2}^{+\infty}\alpha_k\lambda^k/k!$, $\Psi$
reduces to a flow in the canonical form.\\

\noindent
{\bf Proof:} Let us first show that $\xi_{(k)}=\alpha_k\xi_{(1)}$,
$\forall\,k\geq 2$, is a sufficient condition for $\Psi$ to be a
group.  Since Lemma 1 implies
$\phi_\sigma^{(k)}=\phi_{\alpha_k\sigma}^{(1)}$, we have
$\Psi_\lambda=\cdots\circ\phi_{\alpha_k\lambda^k/k!}^{(1)}\circ\cdots
\circ\phi_\lambda^{(1)}=\phi^{(1)}_{\bar{\lambda}}$.  Thus, (i)
$\Psi_\sigma\circ\Psi_\lambda=
\phi^{(1)}_{\bar{\sigma}}\circ\phi^{(1)}_{\bar{\lambda}}=
\phi^{(1)}_{\bar{\sigma}+\bar{\lambda}}=
\phi^{(1)}_{\bar{\tau}}=\Psi_\tau$,
with $\tau=f^{-1}(\bar{\sigma}+\bar{\lambda})$, and (ii)
$\Psi_\lambda^{-1}=\phi^{(1)-1}_{\bar{\lambda}}=
\phi^{(1)}_{-\bar{\lambda}}=\Psi_\rho$, with
$\rho=f^{-1}(-\bar{\lambda})$.

To prove the reverse implication, let us consider first the
case of a knight diffeomorphism of rank two.  If $\Psi$ has to form a 
group, then for any $\lambda,\sigma\in{\rm I\!R}$, there must exist a
$\tau\in{\rm I\!R}$ such that
 $\Psi_\sigma\circ\Psi_\lambda=\Psi_\tau$, i.e.,
\begin{equation}
\phi^{(2)}_{\sigma^2/2}\circ\phi^{(1)}_{\sigma}\circ 
\phi^{(2)}_{\lambda^2/2}\circ\phi^{(1)}_{\lambda}=
\phi^{(2)}_{\tau^2/2}\circ\phi^{(1)}_{\tau}\;.
\label{O}
\end{equation}
Applying Lemma 1 to (\ref{O}) we get, to second order in the parameters
and for an arbitrary tensor $T$,
\begin{eqnarray}
\left(\tau\right. -& \left.\lambda-\sigma\right)\pounds_{\xi_{(1)}}T+
 {1\over 2}\,\left(\tau^2+\lambda^2-2\lambda\tau-\sigma^2\right) 
\pounds_{\xi_{(1)}}^2T \nonumber\\
&+ {1\over 2}\,\left(\tau^2-\lambda^2-\sigma^2\right) 
\pounds_{\xi_{(2)}}T+\cdots=0\;.
\label{00}
\end{eqnarray}
In the limit $\lambda,\sigma\rightarrow 0$, one gets from (\ref{00})
that $\tau=\lambda+\sigma-\alpha_2\lambda\sigma- 
\beta_2\lambda^2-\gamma_2\sigma^2+\cdots\,$, where 
$\alpha_2,\beta_2,\gamma_2$ are unspecified numerical coefficients.  
Substituting back into (\ref{00}) we have that
\begin{equation}
\lambda\sigma\left(\pounds_{\xi_{(2)}}-\alpha_2\pounds_{\xi_{(1)}}\right)T-
\beta_2\lambda^2\pounds_{\xi_{(1)}}T-\gamma_2\sigma^2\pounds_{\xi_{(1)}}T+
\cdots=0\;.
\label{?}
\end{equation}
It is clear that (\ref{?}) can be satisfied only if
$\beta_2=\gamma_2=0$ and 
$\xi_{(2)}=\alpha_2\xi_{(1)}$.  Similarly, considering higher rank
knight diffeomorphisms, one can show that
$\xi_{(k)}=\alpha_k\xi_{(1)}$, $\forall k$.\hfill$\Box$\\

The failure of $\Psi$ to form a group is also related to the following
circumstance.
For any $p\in{\cal M}$, one can define a curve $u_p:{\rm I\!R}\to{\cal
M}$ by
$u_p(\lambda):=\Psi_\lambda(p)$.  However,  these
curves do not form a congruence on $\cal M$.  
For the point $u_p(\lambda)$, say, 
belongs to the image of the curve $u_p$, but also to the one of
$u_{u_p(\lambda)}$, which differs from $u_p$ when at least one of the
$\xi_{(k)}$ is not collinear with $\xi_{(1)}$, since
$u_{u_p(\lambda)}(\sigma)=\Psi_\sigma\circ\Psi_\lambda(p)\neq
\Psi_{\lambda+\sigma}(p)=u_p(\lambda+\sigma)$. 
 Thus, the fundamental property of
a congruence, that each point of $\cal M$ lies
 on the image of one, and only one,
curve, is violated.

Finally, it is perhaps worth pointing out that the result of Lemma 2
{\em cannot\/} be written, even formally, as
\begin{equation}
\Psi_{\lambda}^* T= \sum^{+\infty}_{k=0} \,\frac{\lambda^k}{k!}\,
\pounds^k_{\xi(\lambda)} T
\label{diopoi}
\end{equation}
with 
\begin{equation}
\xi(\lambda):=\sum_{h=0}^{+\infty}{\lambda^h\over h!}\,\xi_{(h+1)}\;,
\end{equation}
because this expression fails to agree with (\ref{lemma2}) for $k\geq 3$. 
One might try to define a vector $\eta(\lambda)$ for which an analog of
(\ref{diopoi}) holds, but this does not seem very useful or
illuminating.

%%%%%%%%%%%%%%%%%%%%%%%%%%%%%%%%%%%%%%%%%%%%%%%%%%%%%%%%%%%%%%%%%%%%%%
%%%%%%%%%%%%%%%%%%%%%%%%%%%%%%%%%%%%%%%%%%%%%%%%%%%%%%%%%%%%%%%%%%%%%%

\subsection{General case}

Knight diffeomorphisms are of a very peculiar form, and the previous
results seem therefore of limited applicability.  This is, however,
not the case, because any one-parameter family of diffeomorphisms can
always be regarded as a one-parameter family of knight diffeomorphisms
--- of infinite rank, in general --- as shown by the following\\

\noindent {\bf Theorem 2:}  Let $\Psi:{\rm I\!R}\times{\cal M}\to {\cal
M}$ be a one-parameter family of diffeomorphisms.  Then $\exists$
$\phi^{(1)},\ldots,\phi^{(k)},\ldots,$ one-parameter groups of
diffeomorphisms of $\cal M$, such that
\begin{equation}
\Psi_\lambda=\cdots\circ\phi^{(k)}_{\lambda^k/k!}\circ\cdots
\circ\phi^{(2)}_{\lambda^2/2}\circ\phi^{(1)}_\lambda\;.
\label{theorem}
\end{equation}

\noindent {\bf Proof:} Consider the action of $\Psi_\lambda$ on a
function $f:{\cal M}\to{\rm I\!R}$.  A Taylor expansion of
$\Psi^*_\lambda f$ gives
\begin{equation} 
\Psi_{\lambda}^* f=\sum_{k=0}^{+\infty}{\lambda^k\over
 k!}\,\left.{{\rm
d}^k~\over{\rm d}\lambda^k}\right|_0\Psi^*_{\lambda} f\;.
\label{ostia}
\end{equation}
The differential operator ${\cal L}_{(1)}$ defined by 
\begin{equation} {\cal L}_{(1)} f:=\left.{{\rm d}~\over{\rm
d}\lambda}\right|_0\Psi_{\lambda}^* f
\end{equation}
is clearly a derivative, so we can define a vector $\xi_{(1)}$ through
$\pounds_{\xi_{(1)}}f:={\cal L}_{(1)}f$.  Similarly,
\begin{equation} {\cal L}_{(2)}f:=\left.{{\rm d}^2~\over{\rm d}
\lambda^2}\right|_0\Psi_{\lambda}^* f-\pounds^2_{\xi_{(1)}}f
\end{equation}
is also a derivative, as one can easily check.  Thus, we define the
 vector $\xi_{(2)}$ such that $\pounds_{\xi_{(2)}}f:={\cal L}_{(2)}f$,
 and so on at higher orders.  Hence, we recover (\ref{lemma2}) for an
 arbitrary $f$.  But if $\varphi$ and $\psi$ are two diffeomorphisms
 on $\cal M$ such that $\varphi^*f=\psi^*f$ for every $f$, it follows
 that $\varphi\equiv\psi$, as it is easy to see.  Thus, we establish
 (\ref{theorem}).\hfill$\Box$\\

It must be noticed that, although we have supposed so far that maps
and fields are analytic, it is possible to give versions of Lemmas 1
and 2, and of Theorem 2, that hold  in the case of  $C^r$ objects
\cite{bi:sm}.  The main change is the substitution of Taylor series
like the one in (\ref{lemma1}) by a finite sum of $n-1$ terms plus a
remainder \cite{bi:ycm}.  The meaning of Theorem 2 is then that any
one-parameter family of diffeomorphisms can be approximated by a
family of knight diffeomorphisms of suitable rank.

%%%%%%%%%%%%%%%%%%%%%%%%%%%%%%%%%%%%%%%%%%%%%%%%%%%%%%%%%%%%%%%%%%%%%%
%%%%%%%%%%%%%%%%%%%%%%%%%%%%%%%%%%%%%%%%%%%%%%%%%%%%%%%%%%%%%%%%%%%%%%

\subsection{Interpreting the literature:  What is what}
\label{secIId}

The abstract mathematical notation that we have used so far is the
most appropriate one for the study of gauge transformations in
perturbation theory from a general point of view.  In order to make
explicit calculations in special cases of physical interest, however,
one must introduce a chart.  Most of the literature on the subject,
therefore, is written in the language of coordinates.  For this
reason, we indicate here how to establish a correspondence between the
two formalisms.  For the sake of simplicity in the notation, we shall
restrict ourselves to consider the action of the pull-back on a vector
field, the extension to one-forms and tensors of higher rank being
straightforward.

Let therefore $({\cal U},x)$ be a chart of $\cal M$, with ${\cal
U}\subseteq{\cal M}$ an open set and $x:{\cal U}\to{\rm I\!R}^m$ given
by $x:p\mapsto (x^0(p),x^1(p),\ldots,x^{m-1}(p))$, $\forall\,p\in{\cal
U}$.  
Since the function $x^\mu: {\cal M}\to{\rm I\!R}$ is differentiable,
we can define the linear map $x^\mu_*$, that
associates to a vector on $\cal M$ its $\mu$-th component in the
coordinate basis defined by the chart $({\cal U},x)$.\footnote{Notice
that, in the language of differential forms,  $x^\mu_*={\rm d}x^\mu$.}

Consider now a vector field $Z$ and its 
 pull-back $\tilde{Z}:=\Psi_{\lambda}^* Z$, which for every
value of $\lambda$ is a new vector field on $\cal M$.  For each point
$p\in{\cal U}$, the components of
$\tilde{Z}(p)$ in the chart $({\cal U},x)$ are
\begin{equation}
\tilde{Z}^\mu(x(p))=\left(x^\mu_* \tilde{Z}\right)(x(p))\;,
\end{equation}
which becomes, using (\ref{phi^*}) and the definition of $\tilde{Z}$,
\begin{equation}
\fl \tilde{Z}^\mu(x(p))=x^\mu_*\left(\tilde{Z}(p)\right)=
x^\mu_*\left(\left(\Psi_\lambda^* Z\right)(p)\right)=
\left(x^\mu\circ\Psi_\lambda^{-1}\right)_*
\left(Z\left(\Psi_\lambda(p)\right)\right)\;.
\end{equation}

Now, let us define a new chart $(\Psi_\lambda({\cal U}),y)$, with
$y^\mu:=x^\mu\circ\Psi_\lambda^{-1}$.  In this way, the
$y$-coordinates of the point $q:=\Psi_\lambda(p)$ coincide with the
$x$-coordinates of the point $p$ from which $q$ has come under the
action of the diffeomorphism: $y^\mu(q)=x^\mu(p)$.  We have then
\begin{equation}
\tilde{Z}^\mu(x(p))=y^\mu_*\left(Z\left(\Psi_\lambda(p)\right)\right)=
\left(y^\mu_*
Z\right)\left(y\left(\Psi_\lambda(p)\right)\right)=\left(y^\mu_*
Z\right)(y(q))\;.
\label{Z}
\end{equation}

Denoting by $Z^\mu$ and $Z'^\mu$ the components of $Z$ in the charts
$({\cal U},x)$ and $(\Psi_\lambda({\cal U}),y)$, respectively, we can
then write
\begin{equation}
\tilde{Z}^\mu(x(p))=Z'^\mu(y(q))=
\left[{\partial y^\mu\over \partial x^\nu}\right]_{x(q)} 
Z^\nu(x(q))\;.
\label{coord-vector}
\end{equation}
That is, the pull-back of $Z$ is characterized by having, in the chart
$({\cal U},x)$ at point $p$, the same components that the original
vector field has in the chart $(\Psi_\lambda({\cal U}),y)$ at point
$q=\Psi_\lambda(p)$.  Or, since $y(q)=x(p)$, one can just write
$\tilde{Z}^\mu(x)=Z'^\mu(x)$, where $x$ simply stands for a point of
${\rm I\!R}^m$.
This property can also be used to {\em
define\/} $\tilde{Z}$, and corresponds to a {\em passive\/}
interpretation of the map $\Psi_\lambda$, regarded as generating a
change in the chart {\em on\/} $\cal M$ rather than a transformation
{\em of\/} $\cal M$ ({\em active\/} view).  Of course, the two
viewpoints are equivalent  \cite{bi:waldbook}, but the active
interpretation is much less confusing.

We end this section by writing down the explicit expression, up to
second order in $\lambda$, for the coordinate transformation
$x\rightarrow y$ associated to a family of diffeomorphisms with
generators $\xi_{(1)},\xi_{(2)},\ldots\,$.  Calling
$q:=\Psi_\lambda(p)$, we have from (\ref{lemma2coord}),
\begin{equation}
\fl x^\mu(q)=x^\mu(p)+\lambda\,\xi_{(1)}^\mu(x(p))+{\lambda^2\over
2}\,\left({\xi_{(1)}^\mu}_{,\nu}(x(p))\ 
\xi_{(1)}^\nu(x(p))+\xi_{(2)}^\mu(x(p))\right)+
\cdots\;.
\label{lemma2coord'}
\end{equation}
By definition, we have also
\begin{eqnarray}
y^\mu(q):=& x^\mu(p)=x^\mu(q)-\lambda\,\xi_{(1)}^\mu(x(p))\nonumber\\
&-{\lambda^2\over 2}\,\left({\xi_{(1)}^\mu}_{,\nu}(x(p))\
 \xi_{(1)}^\nu(x(p))+\xi_{(2)}^\mu(x(p))\right)+\cdots\;.
\label{gosh}
\end{eqnarray}
Expanding the various quantities on the right hand side around $q$, 
(\ref{gosh}) becomes finally
\begin{equation}
\fl y^\mu(q)=x^\mu(q)-\lambda\,\xi_{(1)}^\mu(x(q))+{\lambda^2\over
2}\,\left({\xi_{(1)}^\mu}_{,\nu}(x(q))\ 
\xi_{(1)}^\nu(x(q))-\xi_{(2)}^\mu(x(q))\right)+
\cdots\;.
\label{lemma2coord''}
\end{equation}
Equations (\ref{lemma2coord'}) and (\ref{lemma2coord''}) express the
relationship, in the language of coordinates, between the active and
the passive views.  Whereas (\ref{lemma2coord'}) provides us with
the coordinates, in the {\em same chart\/} $({\cal U},x)$, of the {\em
different points\/} $p$ and $q=\Psi_\lambda(p)$, equation 
(\ref{lemma2coord''}) gives the transformation law between the
coordinates of the {\em same point\/} $q$ in the two {\em different
charts\/} $({\cal U},x)$ and $(\Psi_\lambda({\cal U}),y)$.
An equivalent form of the transformation (\ref{lemma2coord''}) was
already used by Taub in studying the gauge dependence of an
approximate stress energy tensor for gravitational fields \cite{taub}.

Using (\ref{lemma2coord''}) for the actual computation of the
coordinate transformation in (\ref{coord-vector}), and expanding
every term again at second order around $x(p)$, one can derive the
components in the chart $({\cal U},x)$ of the pull-back $\Psi^*_\lambda
Z$ of $Z$, given in terms of a second order expansion formula
involving $Z$ and its partial derivatives along $\xi_{(1)}$ and
$\xi_{(2)}$. Then, properly collecting the various terms, one can
check that this leads to the components of the right hand side of
(\ref{lemma2explic}), i.e., to the components of a Taylor expansion of
$\Psi^*_\lambda Z$ in terms of the Lie derivatives along $\xi_{(1)}$ and
$\xi_{(2)}$ of $Z$.

%%%%%%%%%%%%%%%%%%%%%%%%%%%%%%%%%%%%%%%%%%%%%%%%%%%%%%%%%%%%%%%%%%%%%%
%%%%%%%%%%%%%%%%%%%%%%%%%%%%%%%%%%%%%%%%%%%%%%%%%%%%%%%%%%%%%%%%%%%%%%
%%%%%%%%%%%%%%%%%%%%%%%%%%%%%%%%%%%%%%%%%%%%%%%%%%%%%%%%%%%%%%%%%%%%%%

\section{Perturbations of spacetime and gauge choices}

\setcounter{equation}{0}

In relativistic perturbation theory one tries to find approximate
solutions of the Einstein equation, regarding them as ``small''
deviations from some known exact solution --- the so-called
background.  The perturbation $\Delta T$ in any relevant quantity, say
represented by a tensor field $T$, is defined as the difference
between the value $T$ has in the physical spacetime, and the
background value $T_0$.  However, it is a basic fact of differential
geometry that, in order to make the comparison of tensors meaningful
at all, one has to consider them at the same point.  Since $T$ and
$T_0$ are defined in different spacetimes, they can thus be compared
only after a prescription for identifying points of these spacetimes
is given.  A {\em gauge choice\/} is precisely this, i.e., a map
between the background and the physical spacetime.  Mathematically,
any diffeomorphism between the two spacetimes provides one such
prescription.  A change of this diffeomorphism is then a gauge
transformation, and the freedom one has in choosing it corresponds to
the arbitrariness in the value of the perturbation of $T$ at any given
spacetime point, unless $T$ is gauge-invariant.  This is the essence
of the ``gauge problem,'' which has been discussed in depth in many
papers \cite{bi:sachs,bi:SW,bi:JB,bi:EB,bi:stewart} and review
articles \cite{bi:KS,bi:MFB,bi:RD}.

In order to discuss higher order perturbations and gauge
transformations, and to define gauge invariance, we must formalize the
previous ideas, giving a precise description of what perturbations and
gauge choices are.  Here we shall mainly follow the approach used in
references \cite{bi:SW,bi:stewart,bi:sb,bi:waldbook}  (cf. also
\cite{bi:GL,bi:schutz}). 

Let us thus consider a family of spacetime models $\{({\cal
M},g_\lambda,\tau_\lambda)\}$, where the metric $g_\lambda$ and the
matter fields (here collectively referred to as $\tau_\lambda$)
satisfy the field equation
\begin{equation}
\label{eq:ee} 
{\cal E}\left[g_\lambda,\tau_\lambda\right]=0\;,
\end{equation}
and $\lambda\in{\rm I\!R}$.  We shall assume that $g_\lambda$ and
$\tau_\lambda$ depend smoothly on the parameter $\lambda$, so that
$\lambda$ itself is a measure of the amount by which a specific
$({\cal M},g_\lambda,\tau_\lambda)$ differs from the idealized
background solution $({\cal M},g_0,\tau_0)$, which is supposed to be
known.  In some applications, $\lambda$ is a dimensionless parameter
naturally arising from the physical problem one is dealing with. In
this case one expects the perturbative solution to accurately
approximate the exact one for reasonably small $\lambda$ (see, e.g.,
\cite{bi:gleiseretal}).  In other problems, $\lambda$ can be
introduced as a purely formal parameter, and in the end, for
convenience, one can thus choose $\lambda=1$ for the physical
spacetime, as we shall do in section 5.

This situation is most naturally described by introducing an
$(m+1)$-dimensional manifold $\cal N$, foliated by submanifolds
diffeomorphic to $\cal M$, so that ${\cal N}={\cal M}\times{\rm
I\!R}$.  We shall label each copy of $\cal M$ by the corresponding
value of the parameter $\lambda$.  The manifold $\cal N$ has a natural
differentiable structure which is the direct product of those of $\cal
M$ and ${\rm I\!R}$.  We can then choose charts in which $x^\mu$
($\mu=0,1,\ldots,m-1$) are coordinates on each leave ${\cal
M}_\lambda$, and $x^m\equiv\lambda$.

Now, if a tensor field $T_\lambda$ is given on each ${\cal
M}_\lambda$, we have that a tensor field $T$ is automatically defined
on ${\cal N}$ by the relation $T(p,\lambda):=T_\lambda(p)$, with
$p\in{\cal M}_\lambda$.\footnote{It is worth noticing that tensor
fields on ${\cal N}$ constructed in this way are ``transverse,'' in
the sense that their $m$-th components in the charts we have defined
vanish identically.}  In particular, on each ${\cal M}_\lambda$ one
has a metric $g_\lambda$ and a set of matter fields $\tau_\lambda$,
satisfying the field equation (\ref{eq:ee}); correspondingly, the
fields $g$ and $\tau$ are defined on $\cal N$.

We want now to define the perturbation in any tensor $T$, therefore we
must find a way to compare $T_\lambda$ with $T_0$.  As already said,
this requires a prescription for identifying points of ${\cal
M}_\lambda$ with those of ${\cal M}_0$.  This is easily accomplished
by assigning a diffeomorphism $\varphi_\lambda:{\cal N}\to{\cal N}$
such that $\left.\varphi_\lambda\right|_{\scriptscriptstyle {\cal
M}_0}:{\cal M}_0\to{\cal M}_\lambda$.  Clearly, $\varphi_\lambda$ can
be regarded as the member of a flow $\varphi$ on $\cal N$,
corresponding to the value $\lambda$ of the group parameter.
Therefore, we could equally well give the vector field $X$ that
generates $\varphi$.  In the chart introduced above, $X^m=1$ but,
except for this condition, $X$ remains arbitrary.  With a slight abuse
of terminology, we shall sometimes refer also to such a vector field
as a {\em gauge\/}.

The perturbation can now be defined simply as
\begin{equation}
\Delta T_\lambda:=\left.\varphi^*_\lambda T\right|_{\scriptscriptstyle
{\cal M}_0}-T_0\;.
\label{DeltaT}
\end{equation}
 The first
term on the right hand side of (\ref{DeltaT}) can be
Taylor-expanded to get
\begin{equation}
\Delta T_\lambda=\sum_{k=1}^{+\infty}{\lambda^k\over k!}\,\delta^kT\;,
\label{DeltaT1}
\end{equation}
where
\begin{equation}
\delta^kT:=\left[{{\rm d}^k\varphi^*_\lambda T\over {\rm
d}\lambda^k}\right]_{\lambda=0,{\cal M}_0}\;.
\label{deltakT}
\end{equation}
Equation (\ref{deltakT}) defines then the $k$-th order perturbation of
$T$.  Notice that $\Delta T_\lambda$ and $\delta^k T$ are defined on
${\cal M}_0$; this formalizes the statement one commonly finds in the
literature, that ``perturbations are fields living in the
background.''  It is important to appreciate that the parameter
$\lambda$ labeling the various spacetime models serves also to perform
the expansion (\ref{DeltaT1}), and determines therefore what one means
by ``perturbations of the $k$-th order.''  However, as we have already
pointed out, there are applications where $\lambda$ is, to a large
extent, arbitrary.  In these cases, the split of $\Delta T_\lambda$
into perturbations of first order, second order, and so on, has no
absolute meaning, because a change of $\lambda$, i.e., a
reparametrization of the family of spacetimes, will mix them up.  What
is invariantly defined, is only the quantity $\Delta T_\lambda$,
whereas the various $\delta^kT$ are meaningful only once a choice of
the parameter has been made.

Now, we are interested in those cases in which (\ref{eq:ee}) is too
difficult to solve exactly, so that one looks for approximate
solutions, to some order $n$.  In fact, we can now obtain much simpler
{\it linear\/} equations from (\ref{eq:ee}). At first order,
differentiating (\ref{eq:ee}) with respect to $\lambda$ and setting
$\lambda$ equal to zero, one obtains \cite{bi:waldbook} a linear
equation for $\delta g$ and $\delta\tau$.  At second order, a second
derivative with respect to $\lambda$ of  (\ref{eq:ee}) at
$\lambda=0$ gives an equation of the type
\begin{equation}
L\left[\delta^2 g,\delta^2 \tau\right]=S\left[ \delta g,
\delta\tau\right]\, ,
\end{equation}
which is linear in the second order perturbations  $\delta^2 g$
and $\delta^2 \tau$, and where the first order perturbations $\delta g$,
$\delta\tau$ now appear as known source terms. This can obviously be
extended to higher orders, giving an iterative procedure to 
calculate $\Delta g_\lambda$ and
$\Delta\tau_\lambda$ --- hence $g_\lambda$ and $\tau_\lambda$ 
 --- to the required accuracy.

%%%%%%%%%%%%%%%%%%%%%%%%%%%%%%%%%%%%%%%%%%%%%%%%%%%%%%%%%%%%%%%%%%%%%%
%%%%%%%%%%%%%%%%%%%%%%%%%%%%%%%%%%%%%%%%%%%%%%%%%%%%%%%%%%%%%%%%%%%%%%
%%%%%%%%%%%%%%%%%%%%%%%%%%%%%%%%%%%%%%%%%%%%%%%%%%%%%%%%%%%%%%%%%%%%%%

\section{Gauge invariance and gauge transformations}

\setcounter{equation}{0}

Let us now suppose that two vector fields $X$ and $Y$ are defined on
${\cal N}$, such that they have $X^m=Y^m=1$ everywhere.
Correspondingly, their integral curves define two flows $\varphi$ and
$\psi$ on $\cal N$, that connect any two leaves of the foliation.
Thus $X$ and $Y$ are everywhere transverse to the ${\cal M}_\lambda$,
and points lying on the same integral curve of either of the two are
to be regarded {\em as the same point\/} within the respective gauge:
$\varphi$ and $\psi$ are both point identification maps, i.e., two
different gauge choices.

The fields $X$ and $Y$ can both be used to pull back a generic tensor
field $T$, and to construct therefore two other tensor fields
$\varphi^*_{\lambda} T$ and $ \psi^*_{\lambda} T$, for any given value
of $\lambda$.  In particular, on ${\cal M}_0$ we now have three tensor
fields, i.e., $T_0$, and
\begin{equation}
\label{eq:txy}
T^X_\lambda := \left.\varphi^*_{\lambda} T\right|_0\, , ~~~
T^Y_\lambda := \left.\psi^*_{\lambda} T\right|_0\, , 
\end{equation}
where, for the sake of simplicity, we 
have denoted the restriction to ${\cal M}_0$ of a
tensor field defined over $\cal N$ simply by the suffix 0.

Since  $X$ and $Y$ represent gauge choices for mapping a 
perturbed manifold ${\cal M}_\lambda$ into the unperturbed one ${\cal
M}_0$, $T^X_\lambda$ and $T^Y_\lambda$ are the representations, in
${\cal M}_0$, of the perturbed tensor according to the two gauges.  We 
can write, using (\ref{DeltaT})--(\ref{deltakT}) and Lemma 1,
\begin{eqnarray} 
T^X_\lambda& =&\sum_{k=0}^{+\infty}\frac{\lambda^k}{k!}\,
\delta^k T^X =
\sum_{k=0}^{+\infty}{\lambda^k\over k!}\,
\left.\pounds^k_X T\right|_0 = T_0 +\Delta^\varphi T_\lambda\,
\;,
\label{4.4} \\
T^Y_\lambda&=&\sum_{k=0}^{+\infty}\frac{\lambda^k}{k!}\,
\delta^k T^Y =
\sum_{k=0}^{+\infty}{\lambda^k\over k!}\,
\left.\pounds^k_Y T\right|_0 = T_0 +\Delta^\psi T_\lambda\;.
\label{4.5}
\end{eqnarray}

%%%%%%%%%%%%%%%%%%%%%%%%%%%%%%%%%%%%%%%%%%%%%%%%%%%%%%%%%%%%%%%%%%%%%%
%%%%%%%%%%%%%%%%%%%%%%%%%%%%%%%%%%%%%%%%%%%%%%%%%%%%%%%%%%%%%%%%%%%%%%

\subsection{Gauge invariance}

If $T^X_\lambda=T^Y_\lambda$, for any pair of gauges $X$ and $Y$, we
say that $T$ is {\em totally gauge-invariant\/}.  This is a very
strong condition, because then (\ref{4.4}) and (\ref{4.5}) imply that
$\delta^kT^X=\delta^kT^Y$, for all gauges $X$ and $Y$ and for any $k$.
In any practical case one is however interested in perturbations to a
fixed order $n$; it is thus convenient to weaken the definition above,
saying that $T$ is {\em gauge-invariant to order\/} $n$ iff
$\delta^kT^X=\delta^kT^Y$ for any two gauges $X$ and $Y$, and $\forall
k\leq n$.  We have then the following ($\delta^0T:=T_0$, $\delta
T:=\delta^1 T$) \\

\noindent {\bf Proposition 1:} A tensor field $T$ is 
gauge-invariant to order $n\geq 1$ iff $\pounds_\xi\delta^kT=0$, for
any vector field $\xi$ on $\cal M$ and $\forall k<n$.\\

\noindent {\bf Proof:}  Let us first show that the statement is true
for $n=1$.  In fact, if $\delta T^X=\delta T^Y$, we have
$\pounds_{X-Y}T|_0=0$.  But since $X$ and $Y$ define arbitrary gauges, it
follows that $X-Y$ is an arbitrary vector field $\xi$ with $\xi^m=0$,
i.e., tangent to $\cal M$.  Let us now suppose that
the statement is true for some $n$.  Then, if one has also 
$\delta^{n+1}T^X|_0=\delta^{n+1}T^Y|_0$, it follows that
$\pounds_{X-Y}\delta^nT^X=0$, and we establish the result by
induction over $n$.\hfill$\Box$\\

As a consequence, $T$ is gauge-invariant to order $n$ iff $T_0$ and
all its perturbations of order lower than $n$ are, in any gauge,
either vanishing, or constant scalars, or a combination of Kronecker
deltas with constant coefficients.  Thus, this generalizes to an
arbitrary order $n$ the results of references
\cite{bi:sachs,bi:SW,bi:stewart,bi:sb}. Further, it then follows that
$T$ is totally gauge-invariant iff it is a combination of Kronecker
deltas with coefficients depending only on $\lambda$.

%%%%%%%%%%%%%%%%%%%%%%%%%%%%%%%%%%%%%%%%%%%%%%%%%%%%%%%%%%%%%%%%%%%%%%
%%%%%%%%%%%%%%%%%%%%%%%%%%%%%%%%%%%%%%%%%%%%%%%%%%%%%%%%%%%%%%%%%%%%%%

\subsection{Gauge transformations}

\begin{figure}
\vspace*{310pt}
\includegraphics{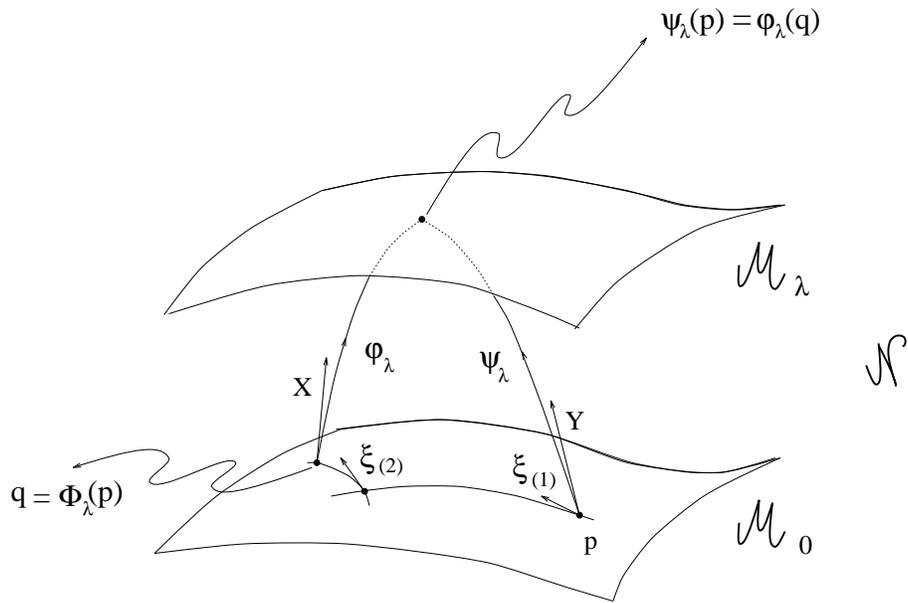}
\caption{The action of a gauge transformation $\Phi_\lambda$,
represented on the background spacetime ${\cal M}_0$ by its second
order approximation, generated by the two vector fields $\xi_{(1)}$ 
and $\xi_{(2)}$.}\label{fig:GT}
\end{figure}

\begin{figure}
\vspace{270pt}
\includegraphics{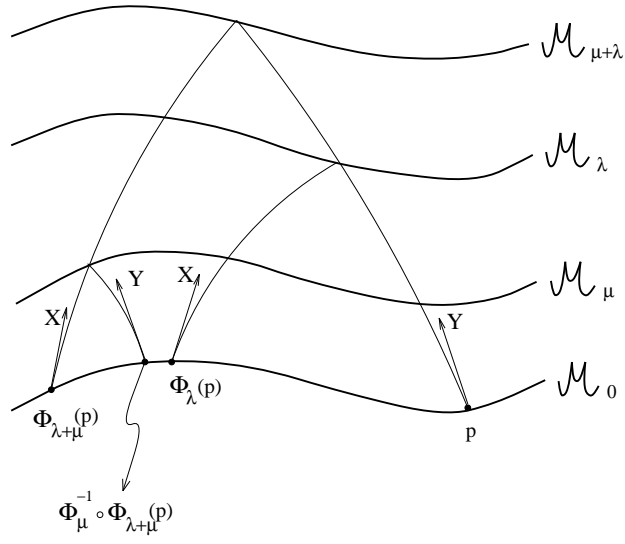}
\caption{If $X$ and $Y$ do not commute, $\Phi$ is not a flow on 
${\cal M}_0$.}\label{fig:commut}
\end{figure}

If a tensor $T$ is not gauge-invariant, it is important to know how
its representation on ${\cal M}_0$ changes under a gauge
transformation.  To this purpose, it is useful to define, for each
value of $\lambda\,\in{\rm I\!R}$, the diffeomorphism
$\Phi_\lambda\,:\,{\cal M}_0\to {\cal M}_0$ given by 
\begin{equation}
\Phi_\lambda :=\varphi_{-\lambda}\circ\psi_{\lambda}\;.
\end{equation}
The action of $\Phi_\lambda$ is illustrated in figure \ref{fig:GT}.
We must stress that $\Phi:{\rm I\!R}\times{\cal
M}_0\to{\cal M}_0$ so defined, {\em is not\/} a one-parameter group of
diffeomorphisms on ${\cal M}_0$.  In fact,
$\Phi_{-\lambda}\not=\Phi^{-1}_\lambda$, and
$\Phi_{\lambda+\sigma} \not= \Phi_\sigma\circ\Phi_\lambda$, essentially
because the fields $X$ and $Y$ have, in general, a non vanishing
commutator, as depicted in figure \ref{fig:commut}.  However, Theorem 2 
guarantees that, to order $n$ in $\lambda$, the one-parameter family of
diffeomorphisms $\Phi$ can always be approximated by a one-parameter
family of knight diffeomorphisms of rank $n$\footnote{This result
confirms a  claim in \cite{bi:GL}.}  
 (see figure \ref{fig:GT}
for the action of $\Phi_\lambda$ to second order).

It is very easy to see that the tensor fields $T^X_\lambda$ and
$T^Y_\lambda$ defined by the gauges $\varphi$ and $\psi$ are connected
by the linear map $\Phi_\lambda^*$:
\begin{equation}
T^Y_\lambda=\left.\psi^*_{\lambda} T\right|_0=
\left.(\psi^*_{\lambda}\varphi^*_{-\lambda}\varphi^*_{\lambda}
T)\right|_0 =\left.\Phi_\lambda^*(\varphi^*_{\lambda}T)\right|_0
=\Phi_\lambda^* T^X_\lambda\;.
\end{equation}
Thus, Theorem 2 allows us to use (\ref{lemma2}) as a generating
formula for a gauge transformation to an arbitrary order $n$:
\begin{eqnarray}
\fl T^Y_\lambda=\sum_{l_1=0}^{+\infty} \sum_{l_2=0}^{+\infty}\cdots
\sum_{l_k=0}^{+\infty}\cdots \, 
{\lambda^{l_1+2l_2+\cdots+kl_k+\cdots}\over
2^{l_2}\cdots (k!)^{l_k}\cdots l_1!l_2!\cdots
l_k!\cdots}\,\pounds^{l_1}_{\xi_{(1)}}\pounds^{l_2}_{\xi_{(2)}}
\cdots\pounds^{l_k}_{\xi_{(k)}}\cdots
 T^X_\lambda\;.\nonumber\\
\label{lemma2applied}
\end{eqnarray}
To third order, we have explicitly 
\begin{eqnarray}
\label{eq:tgt} T^Y_\lambda=&T^X_\lambda + 
\lambda\pounds_{\xi_{(1)}} T^X_\lambda
+\frac{\lambda^2}{2}\,\left(\pounds^2_{\xi_{(1)}}+
\pounds_{\xi_{(2)}}\right) T^X_\lambda\nonumber\\
&+{\lambda^3\over
3!}\,\left(\pounds^3_{\xi_{(1)}}
+3\,\pounds_{\xi_{(1)}}\pounds_{\xi_{(2)}}+
\pounds_{\xi_{(3)}}\right) T^X_\lambda+\ldots\;,
\label{fund}
\end{eqnarray}
where $\xi_{(1)}$ and $\xi_{(2)}$ are now the first two generators of
$\Phi_\lambda$, or of the gauge transformation, if one prefers.

We can now relate the perturbations in the two gauges.  To the
lowest orders, this is easy to do explicitly:\\

\noindent {\bf Proposition 2:} Given a tensor field $T$,  
the relations between the first, second, and third 
order perturbations of $T$ in  two different gauges 
are\footnote{A second order gauge transformation equivalent to
(\ref{second}) has  recently been given in \cite{bi:FW}, see their
section III~C.}:
\begin{equation} 
\delta T^Y-\delta T^X=\pounds_{\xi_{(1)}}T_0\;;
\label{first}
\end{equation} 
\begin{equation}
\delta^2 T^Y-\delta^2 T^X=\left(\pounds_{\xi_{(2)}}+
\pounds^2_{\xi_{(1)}}\right)T_0 
+2\pounds_{\xi_{(1)}}\delta T^X \;;
\label{second}
\end{equation}
\begin{eqnarray}
\delta^3 T^Y-\delta^3
T^X=&\left(\pounds_{\xi_{(3)}}
+3\pounds_{\xi_{(1)}}\pounds_{\xi_{(2)}}+
\pounds^3_{\xi_{(1)}}\right)T_0\nonumber\\
&+3\left(\pounds_{\xi_{(2)}}+
\pounds^2_{\xi_{(1)}}\right)\delta T
+3\pounds_{\xi_{(1)}}\delta^2 T^X
\;.
\label{third}
\end{eqnarray}
\noindent {\bf Proof:}  Substitute (\ref{4.4}) and (\ref{4.5})
into (\ref{eq:tgt}).\hfill$\Box$\\

This result is consistent with Proposition 1, of course.  Equation
(\ref{first}) implies that $T_\lambda$ is gauge-invariant to the first
order iff $\pounds_\xi T_0=0$, for any vector field $\xi$ on $\cal M$.
In particular, one must have $\pounds_{\xi_{(2)}}T_0=0$, and therefore
(\ref{second}) leads to $\pounds_\xi\delta T=0$.  Similarly, one
has then $\pounds_\xi\delta^2T=0$ from (\ref{third}), and so on
recursively at higher orders.

It is also possible to find the explicit expressions, in terms of $X$
and $Y$, for the generators $\xi_{(k)}$ of a gauge transformation:\\

\noindent {\bf Proposition 3:} The first three generators of the
one-parameter family of  diffeomorphisms $\Phi$ are: 
\begin{equation}
\xi_{(1)}=Y-X\;;
\label{xi1}
\end{equation}
\begin{equation}
\xi_{(2)}=[X,Y]\;;
\label{xi2}
\end{equation}
\begin{equation}
\xi_{(3)}=[2X-Y,[X,Y]]\;.
\label{xi3}
\end{equation}

\noindent {\bf Proof:} Substituting (\ref{4.4}) and (\ref{4.5}) into
(\ref{fund}), using the fact that $\pounds_{\xi_{(k)}}\lambda=0$, and
identifying terms of first order in $\lambda$, we find
\begin{equation}
\pounds_{\xi_{(1)}} T_0 = \left. \pounds_{Y-X} T\right|_0\, .
\end{equation}
Since $Y^m-X^m=0$ and $T$ is arbitrary, we have (\ref{xi1}).  
Substituting back, and identifying terms of order
$\lambda^2$, we have now, similarly, 
\begin{equation}
\pounds_{\xi_{(2)}} T_0 =  \left. \pounds_{[X,Y]} T\right|_0\, .
\end{equation}
But $[X,Y]^m=0$, so we obtain (\ref{xi2}).  Analogously, one finds
\begin{equation}
\pounds_{\xi_{(3)}}T_0=\left.\pounds_{[2X-Y,[X,Y]]}T\right|_0\;.
\end{equation}
Since $[2X-Y,[X,Y]]^m=0$, we get (\ref{xi3}).\hfill$\Box$\\

%%%%%%%%%%%%%%%%%%%%%%%%%%%%%%%%%%%%%%%%%%%%%%%%%%%%%%%%%%%%%%%%%%%%%%
%%%%%%%%%%%%%%%%%%%%%%%%%%%%%%%%%%%%%%%%%%%%%%%%%%%%%%%%%%%%%%%%%%%%%%
%%%%%%%%%%%%%%%%%%%%%%%%%%%%%%%%%%%%%%%%%%%%%%%%%%%%%%%%%%%%%%%%%%%%%%

\section{An example from cosmology}

\setcounter{equation}{0}

As an example of the applications of the gauge transformation
obtained, we now show how the perturbations on a spatially flat
Robertson--Walker background in two different gauges are related, up
to second order.  We shall first consider the metric perturbations,
then those in the energy density and 4-velocity of matter. Thus in
this section we choose $m=4$, so that the Greek indices
$\mu,\nu,\ldots$ take values from 0 to 3, and the Latin ones
$i,j,\ldots$ from 1 to 3.

The components of a perturbed spatially flat Robertson--Walker 
metric can be written as 
\begin{equation}\label{eq:m1}
g_{00}=-a(\tau)^2\left(1+2\sum_{r=1}^{+\infty}{1\over
r!}\ \psi^{(r)}\right)\;,
\end{equation}
\begin{equation}\label{eq:m2}
g_{0i}=a(\tau)^2\sum_{r=1}^{+\infty}{1\over
r!}\ \omega^{(r)}_i\;,
\end{equation}
\begin{equation}\label{eq:m3}
g_{ij}=a(\tau)^2\left[\left( 1-2 \sum_{r=1}^{+\infty}{1\over
r!}\ \phi^{(r)}\right)\delta_{ij}+\sum_{r=1}^{+\infty}{1\over
r!}\ \chi^{(r)}_{ij}\right]\;,
\end{equation}
where\footnote{Indices are raised and lowered
using $\delta^{ij}$ and $\delta_{ij}$, respectively.}
$\chi^{(r)i}_{i}=0$, and $\tau$ is
the conformal time.  The functions $\psi^{(r)}$, $\omega^{(r)}_i$,
$\phi^{(r)}$, and $\chi^{(r)}_{ij}$ represent the
 $r$-th order perturbation of the metric.

It is standard to use a non-local splitting of perturbations into the
so-called scalar, vector and tensor parts, where scalar (or
longitudinal) parts are those related to a scalar potential, vector
parts are those related to transverse (divergence-free, or solenoidal)
vector fields, and tensor parts to transverse trace-free tensors.
Such a splitting generalizes the Helmholtz theorem of standard vector
calculus (see, e.g.,  \cite{bi:arfken}), and can be performed on
any spacetime (see, e.g.,  \cite{bi:stewart} and references
therein) imposing suitable boundary conditions.  In our case, the
shift $\omega_i^{(r)}$ can be decomposed as
\begin{equation}
\omega_i^{(r)}= \partial_i\omega^{(r)\|} +\omega_i^{(r)\perp}\;,
\end{equation}
where $\omega_i^{(r)\perp}$ is a solenoidal vector, i.e.,
$\partial^i\omega_i^{(r)\perp}=0$.  
Similarly, the traceless part of the
spatial metric can be decomposed at any order as
\begin{equation}
\chi^{(r)}_{ij}={\rm D}_{ij}\chi^{(r)\|}+\partial_i\chi^{(r)\bot}_j+
\partial_j\chi^{(r)\bot}_i+\chi^{(r)\top}_{ij}\;,
\end{equation}
where $\chi^{(r)\|}$ is a suitable function, $\chi^{(r)\bot}_i$ is a
solenoidal vector field, and 
$\partial^i\chi^{(r)\top}_{ij}=0$; hereafter,
\begin{equation}
{\rm D}_{ij}:=\partial_i\partial_j-
{1\over 3}\,\delta_{ij}\nabla^2\;.
\end{equation}

Now, consider the energy density $\mu$, or any other scalar
 that depends only  on $\tau$ at zero order:
this can be written as
\begin{equation}\label{eq:mu}
\mu=\mu_{(0)}+\sum_{r=1}^{+\infty}{1\over
r!}\ \delta^r \mu\; .
\end{equation}
For the 4-velocity $u^\mu$ of matter we can write
\begin{equation}\label{eq:4v}
u^\mu=\frac{1}{a}\left(\delta^\mu_0 +\sum_{r=1}^{+\infty}{1\over
r!}\ v^\mu_{(r)}\right)\; .
\end{equation}
In addition, $u^\mu$ is subject to the normalization condition $u^\mu
 u_\mu=-1$; therefore at any order the time component $v^0_{(r)}$ is
 related to the lapse perturbation, $\psi_{(r)}$.  For the first and
 second order perturbations we obtain, in any gauge:
\begin{eqnarray}
\label{eq:v0psi1}
v^0_{(1)} &= &-\psi_{(1)}\; ; \\
\label{eq:v0psi2}
v^0_{(2)} &= &-\psi_{(2)}+3\psi^2_{(1)}+2\omega^{(1)}_i v^i_{(1)} 
+v^{(1)}_iv_{(1)}^i\; .
\end{eqnarray}
The velocity perturbation $v^i_{(r)}$ can also be split into a scalar
and vector (solenoidal) part:
\begin{equation}
v^i_{(r)}=\partial^i v_{(r)}^{\|} +v^i_{(r)\perp}\;.
\end{equation} 

As we have seen in the last section, the gauge transformation is
determined by the vectors $\xi_{(r)}$.  Splitting their time and space
parts, one can write
\begin{equation}
\xi_{(r)}^0=\alpha^{(r)}\;,
\end{equation}
and
\begin{equation}
\xi_{(r)}^i=\partial^i\beta^{(r)}+d^{(r)i}\;,
\end{equation}
with $\partial_i d^{(r)i}=0$.

%%%%%%%%%%%%%%%%%%%%%%%%%%%%%%%%%%%%%%%%%%%%%%%%%%%%%%%%%%%%%%%%%%%%%%
%%%%%%%%%%%%%%%%%%%%%%%%%%%%%%%%%%%%%%%%%%%%%%%%%%%%%%%%%%%%%%%%%%%%%%

\subsection{First order}

We begin by reviewing briefly some well-known results about first
order gauge transformations, as we shall need them in the following.  
From now on, we will drop the suffixes $X$ and $Y$ used previously to
denote the ``old'' and ``new'' gauge choices, simply using a tilde to
denote quantities in the new gauge.

For the sake of completeness, we recall here  the basic coordinate
expressions of the Lie derivative along a vector field $\xi$. 
For a  scalar $f$, a vector $Z$ and a covariant tensor
$T$ of rank two, these are, respectively:
\begin{eqnarray}\label{eq:liedev1}
\pounds_\xi f & = &  f_{,\mu}\xi^\mu\; ; 
\\
\label{eq:liedev2}
\pounds_\xi Z^\mu & = & Z^\mu_{,\nu}\xi^\nu -\xi^\mu_{,\nu}Z^\nu\; ;
\\
\label{eq:liedev3}
\pounds_\xi T_{\mu\nu} & = & T_{\mu\nu,\sigma}\xi^\sigma
+\xi^\sigma_{,\mu} T_{\sigma\nu}+\xi^\sigma_{,\nu} T_{\mu\sigma}\; .
\end{eqnarray}
Expressions for any other tensor can easily be derived from these.

%%%%%%%%%%%%%%%%%%%%%%%%%%%%%%%%%%%%%%%%%%%%%%%%%%%%%%%%%%%%%%%%%%%%%%

\subsubsection{General transformation}

From  (\ref{first}), it follows that the first order perturbations
of the metric transform as
\begin{equation}
\delta\tilde{g}_{\mu\nu}=\delta
g_{\mu\nu}+\pounds_{\xi_{(1)}}g^{(0)}_{\mu\nu}\;,
\end{equation}
where $g^{(0)}_{\mu\nu}$ is the background metric.  Therefore, using
(\ref{eq:liedev3}), we obtain the following transformations for
the first order quantities appearing in (\ref{eq:m1})--(\ref{eq:m3}):
\begin{equation}
\tilde{\psi}_{(1)}=\psi_{(1)}+\alpha_{(1)}^{\prime}
+{a'\over a}\,\alpha_{(1)}\;;
\label{pertpsi}
\end{equation}
\begin{equation}
\tilde{\omega}^{(1)}_i=\omega^{(1)}_i-\alpha^{(1)}_{,i}+
\beta^{(1)\prime}_{,i}+d^{(1)\prime}_i\;;
\end{equation}
\begin{equation}
\tilde{\phi}_{(1)}=\phi_{(1)}-
{1\over 3}\,\nabla^2\beta_{(1)}-{a'\over
a}\,\alpha_{(1)}\;;
\end{equation}
\begin{equation}
\tilde{\chi}^{(1)}_{ij}=\chi^{(1)}_{ij}+
2{\rm D}_{ij}\beta^{(1)}+d^{(1)}_{i,j}+d^{(1)}_{j,i}\;;
\label{pertchi}
\end{equation}
where a prime denotes the derivative with respect to $\tau$.

For a scalar $\mu$, from (\ref{first}), (\ref{eq:mu}), and 
(\ref{eq:liedev1}) we have
\begin{equation}\label{eq:mut}
\delta\tilde{\mu}=\delta\mu +\mu_{(0)}^\prime\alpha_{(1)}\; .
\end{equation}
For the 4-velocity $u^\mu$, we have from (\ref{first}) 
\begin{equation}
\delta \tilde{u}^\mu=\delta u^\mu 
+\pounds_{\xi_{(1)}}u^\mu_{(0)}\;.
\end{equation} 
Using (\ref{eq:liedev2}) and (\ref{eq:4v}) this gives:
\begin{eqnarray}
\label{eq:v0}
\tilde{v}^0_{(1)} & = & v^0_{(1)}
-\frac{a'}{a}\alpha_{(1)}-\alpha_{(1)}^\prime\; ; \\
\label{eq:vi}
\tilde{v}^i_{(1)} & = & v^i_{(1)}
-\beta_{(1)}^{\prime,i}-d_{(1)}^{i\prime}\; .
\end{eqnarray}

The 4-velocity is however subject to the constraint (\ref{eq:v0psi1}),
therefore (\ref{eq:v0}) reduces to (\ref{pertpsi}).

%%%%%%%%%%%%%%%%%%%%%%%%%%%%%%%%%%%%%%%%%%%%%%%%%%%%%%%%%%%%%%%%%%%%%%

\subsubsection{Transforming from 
the synchronous to the Poisson gauge}

Let us now consider the particular case of the transformation from the
synchronous to the Poisson gauge.  The {\em synchronous gauge\/} has
been the one most frequently used in cosmological perturbation theory;
it is defined by the conditions $g_{00}=-a(\tau)^2$ and $g_{0i}=0$
 \cite{bi:li46}. In this way the four degrees of freedom associated
with the coordinate (or diffeomorphism) invariance of the theory are
fixed.  The {\em Poisson gauge\/}, recently discussed by Bertschinger
 \cite{bi:bertschinger}, is instead defined by
${\omega_i}^{(r),i}={\chi_{ij}}^{(r),j}=0$.  Then, one scalar degree
of freedom is eliminated from $g_{0i}$ ($\omega^{(r)\|}=0$), and one
scalar and two vector degrees of freedom from $g_{ij}$
($\chi^{(r)\|}=\chi_i^{(r)\bot}=0$).  This gauge generalizes the
well-known {\em longitudinal gauge\/} to include vector and tensor
modes.  This gauge, in which $\omega_i^{(r)}=\chi_{ij}^{(r)}=0$, has
been widely used in the literature to investigate the evolution of
scalar perturbations  \cite{bi:MFB}. Since the vector and tensor modes
are set to zero by hand, the longitudinal gauge cannot be used to
study perturbations beyond the linear regime, because in the nonlinear
case the scalar, vector, and tensor modes are dynamically
coupled.\footnote{In other words, even if one starts with purely
scalar linear perturbations as initial conditions for the second order
theory, vector and tensor modes are dynamically
generated \cite{bi:MPS}.}

Given the perturbation of the metric in one gauge, it is easy to
obtain, from (\ref{pertpsi})--(\ref{pertchi}), the gauge
transformation to the other one, hence the perturbations in the new
gauge.  In the particular case of the synchronous and Poisson gauges,
we have:
\begin{equation}
\psi_{\scriptscriptstyle \rm P}^{(1)}
=\alpha^{(1)\prime}+{a'\over a}\,\alpha^{(1)}\;;
\label{i}
\end{equation}
\begin{equation}
\alpha^{(1)}=\beta^{(1)\prime}\;;
\label{iii}
\end{equation}
\begin{equation}
\omega_{{\scriptscriptstyle\rm P}\ i}^{(1)}=d^{(1)\prime}_i\;;
\label{iv}
\end{equation}
\begin{equation}
\phi_{\scriptscriptstyle \rm P}^{(1)}=\phi_{\scriptscriptstyle
\rm S}^{(1)}-{1\over 3}\,\nabla^2\beta^{(1)}-
{a'\over a}\,\alpha^{(1)}\;;
\label{ii}
\end{equation}
\begin{equation}
{\rm D}_{ij}\left(\chi_{\scriptscriptstyle
\rm S}^{(1)\|}+2\beta^{(1)}\right)=0\;;
\label{v}
\end{equation}
\begin{equation}
\chi_{{\scriptscriptstyle
\rm S}\ (i,j)}^{(1)\bot}+d^{(1)}_{(i,j)}=0\;;
\label{vi}
\end{equation}
\begin{equation}
\chi_{{\scriptscriptstyle\rm  P}\ ij}^{(1)\top}=
\chi_{{\scriptscriptstyle
\rm S}\ ij}^{(1)\top}\;.
\label{vii}
\end{equation}

The parameters $\alpha^{(1)}$, $\beta^{(1)}$, and $d^{(1)}_i$ of the
gauge transformation can be obtained from (\ref{iii}),
(\ref{v}), and (\ref{vi}) respectively, while the transformed metric
perturbations follow from (\ref{i}), (\ref{iv}), (\ref{ii}), and
(\ref{vii}).

Once these parameters are known, the transformation rules for the
energy density $\mu$ or any other scalar, and those for the 4-velocity
$u^\mu$, follow trivially from (\ref{eq:mut}), (\ref{eq:v0}),
and (\ref{eq:vi}).

%%%%%%%%%%%%%%%%%%%%%%%%%%%%%%%%%%%%%%%%%%%%%%%%%%%%%%%%%%%%%%%%%%%%%%
%%%%%%%%%%%%%%%%%%%%%%%%%%%%%%%%%%%%%%%%%%%%%%%%%%%%%%%%%%%%%%%%%%%%%%

\subsection{Second order}

%%%%%%%%%%%%%%%%%%%%%%%%%%%%%%%%%%%%%%%%%%%%%%%%%%%%%%%%%%%%%%%%%%%%%%

We now extend these well-known transformation rules of linear
metric perturbations to the second order.

\subsubsection{General transformation}

Second order perturbations of the metric transform, according to 
(\ref{second}), as
\begin{equation}
\delta^2\tilde{g}_{\mu\nu}=\delta^2g_{\mu\nu}
+2\pounds_{\xi_{(1)}}\delta
g_{\mu\nu}+\pounds^2_{\xi_{(1)}}g^{(0)}_{\mu\nu}+
\pounds_{\xi_{(2)}}g^{(0)}_{\mu\nu}\;.
\end{equation}
This leads to the following transformations in the second order
quantities appearing in (\ref{eq:m1})--(\ref{eq:m3}):\\

\noindent
%%%%%%%%%%%%%%%%%%%%%%%%%%%%%%%%%%%%%%%%%%%%%%%%%%%%%%%%%%%%%%%%%%%
{\bf lapse perturbation}
%%%%%%%%%%%%%%%%%%%%%%%%%%%%%%%%%%%%%%%%%%%%%%%%%%%%%%%%%%%%%%%%%%%
\begin{eqnarray}\label{eq:psi2}
\fl \tilde{\psi}^{(2)} = &\psi^{(2)}
+\alpha^{(1)}\left[ 2\left(\psi_{(1)}^{\prime} 
+ 2\frac{a'}{a}\psi_{(1)}\right)
+ \alpha_{(1)}^{\prime\prime}  
+5\frac{a'}{a}\alpha_{(1)}^{\prime}
+\left(\frac{a''}{a} 
+\frac{a^{\prime 2}}{a^2}\right) \alpha_{(1)}\right] 
\nonumber \\
\fl & +\xi_{(1)}^i\left(2\psi^{(1)}_{,i} + \alpha^{(1)\prime}_{,i}
+\frac{a'}{a}\alpha^{(1)}_{,i}\right)
+2\alpha_{(1)}^{\prime}\left(2\psi_{(1)} 
+\alpha_{(1)}^{\prime}\right) 
\\  \fl &  
+\xi^{i\prime}_{(1)}\left(\alpha^{(1)}_{,i} -\xi^{(1)\prime}_{i}
-2\omega^{(1)}_i \right)
+\alpha_{(2)}^{\prime}+\frac{a'}{a}\alpha_{(2)}\; ;\nonumber
\end{eqnarray}
%%%%%%%%%%%%%%%%%%%%%%%%%%%%%%%%%%%%%%%%%%%%%%%%%%%%%%%%%%%%%%
{\bf shift perturbation}
%%%%%%%%%%%%%%%%%%%%%%%%%%%%%%%%%%%%%%%%%%%%%%%%%%%%%%%%%%%%%%
\begin{eqnarray}\label{eq:omega2}
\fl \tilde{\omega}^{(2)}_i & = & \omega^{(2)}_i
-4\psi^{(1)}\alpha^{(1)}_{,i}
+\alpha^{(1)}\left[
2\left(\omega^{(1)\prime}_i +2\frac{a'}{a}\omega^{(1)}_i\right)
-\alpha^{(1)\prime}_{,i} + \xi^{(1)\prime\prime}_{i}\right.
\nonumber\\
\fl && - \left. 4\frac{a'}{a}\left(\alpha^{(1)}_{,i}
 -\xi^{(1)\prime}_i\right)\right]+ \xi^j_{(1)}
\left(2\omega^{(1)}_{i,j}-\alpha^{(1)}_{,ij}
+\xi^{(1)\prime}_{i,j}\right) \\
\fl &&+\alpha_{(1)}^{\prime}\left( 2\omega^{(1)}_i
-3\alpha^{(1)}_{,i}  
+\xi^{(1)\prime}_{i}\right)
+ \xi^{j\prime}_{(1)}\left(-4\phi^{(1)}\delta_{ij}+2\chi^{(1)}_{ij}
+2\xi^{(1)}_{j,i}+\xi^{(1)}_{i,j}\right)\nonumber\\
\fl &&
+\xi^j_{(1),i}\left(2\omega^{(1)}_j-\alpha^{(1)}_{,j}\right)
-\alpha^{(2)}_{,i} +\xi^{(2)\prime}_{i}\; ;\nonumber
\end{eqnarray}
%%%%%%%%%%%%%%%%%%%%%%%%%%%%%%%%%%%%%%%%%%%%%%%%%%%%%%%%%%%%%%%%%%%%%%
{\bf spatial metric, trace}
%%%%%%%%%%%%%%%%%%%%%%%%%%%%%%%%%%%%%%%%%%%%%%%%%%%%%%%%%%%%%%%%%%%%%%
\begin{eqnarray}\label{eq:phi2}
\fl \tilde{\phi}^{(2)} & = & \phi^{(2)}
+\alpha^{(1)}\left[
2\left(\phi_{(1)}^{\prime}+
2\frac{a'}{a}\phi_{(1)}\right)
-\left(\frac{a''}{a}+\frac{a^{\prime 2}}{a^2}\right)\alpha_{(1)}
-\frac{a'}{a} \alpha_{(1)}^{\prime}\right]\nonumber\\
\fl &&+\xi^i_{(1)}\left( 2\phi^{(1)}_{,i}-
\frac{a'}{a}\alpha^{(1)}_{,i}\right)
-\frac{1}{3}\left(-4\phi_{(1)}+\alpha_{(1)}\partial_0
+\xi^i_{(1)}\partial_i
+4\frac{a'}{a}\alpha_{(1)}\right)\nabla^2\beta_{(1)}\nonumber\\
\fl &&-\frac{1}{3}\left(2\omega^i_{(1)}-\alpha^{,i}_{(1)}
+\xi^{i\prime}_{(1)}\right)\alpha^{(1)}_{,i} 
-\frac{1}{3}\left(2\chi^{(1)}_{ij}+\xi^{(1)}_{i,j}+
\xi^{(1)}_{j,i}\right)
\xi^{j,i}_{(1)}\\
\fl &&-\frac{a'}{a}\alpha_{(2)} -\frac{1}{3}\nabla^2\beta_{(2)}\;
 ;\nonumber
\end{eqnarray}
%%%%%%%%%%%%%%%%%%%%%%%%%%%%%%%%%%%%%%%%%%%%%%%%%%%%%%%%%%%%%
{\bf spatial metric, traceless part}
%%%%%%%%%%%%%%%%%%%%%%%%%%%%%%%%%%%%%%%%%%%%%%%%%%%%%%%%%%%%%
\begin{eqnarray}\label{eq:chi2}
\fl \tilde{\chi}^{(2)}_{ij} & = &\chi^{(2)}_{ij}
+2\left(\chi^{(1)\prime}_{ij}
+2\frac{a'}{a}\chi^{(1)}_{ij}\right)\alpha_{(1)}
+2\chi^{(1)}_{ij,k}\xi^k_{(1)} 
\nonumber \\
\fl &+& 2\left(-4\phi_{(1)}+\alpha_{(1)}\partial_0
+\xi^k_{(1)}\partial_k+4\frac{a'}{a}\alpha_{(1)}\right)
\left(d^{(1)}_{(i,j)}+{\rm D}_{ij}\beta_{(1)}\right) 
 \nonumber \\
\fl &+& 2\left[\left(2\omega^{(1)}_{(i}-\alpha^{(1)}_{,(i}
+\xi^{(1)\prime}_{(i}\right) \alpha^{(1)}_{,j)}
-\frac{1}{3}\delta_{ij}
\left(2\omega^k_{(1)}-\alpha^{,k}_{(1)}
+\xi^{k\prime}_{(1)}\right)\alpha^{(1)}_{,k}\right] \\
\fl &+& 2\left[\left(2\chi^{(1)}_{(i|k|}
+\xi^{(1)}_{k,(i}+\xi^{(1)}_{(i,|k|}\right) \xi^{(1)k}_{,j)}
-\frac{1}{3}\delta_{ij}\left(2\chi^{(1)}_{lk}
+\xi^{(1)}_{k,l}+\xi^{(1)}_{l,k}\right) \xi_{(1)}^{k,l}\right]
 \nonumber \\
\fl &+& 2\left(d^{(2)}_{(i,j)} +{\rm D}_{ij}\beta_{(2)}\right)\;.
\nonumber
\end{eqnarray}

For the energy density $\mu$, or any other scalar, we have from 
(\ref{second}):
\begin{equation}
\delta^2\tilde{\mu}=\delta^2\mu +\left(\pounds_{\xi_{(2)}} +
\pounds^2_{\xi_{(1)}}\right)\mu_{(0)}
+2\pounds_{\xi_{(1)}}\delta\mu\;.
\end{equation}
From this we obtain, using (\ref{eq:liedev1}):
\begin{eqnarray}\label{eq:mut2}
\delta^2\tilde{\mu}=&\delta^2\mu +\mu_{(0)}^\prime\alpha_{(2)} +
\alpha_{(1)}\left(\mu_{(0)}^{\prime\prime}\alpha_{(1)}
+\mu_{(0)}^\prime\alpha_{(1)}^\prime+2\delta\mu^\prime\right)
\nonumber\\
&+\xi^i_{(1)}\left(\mu^\prime_{(0)}\alpha^{(1)}_{,i}
+2\delta\mu_{,i}\right).
\end{eqnarray}.

For the 4-velocity $u^\mu$,  we have from
(\ref{second}): 
\begin{equation}
\delta^2 \tilde{u}^\mu =\delta^2u^\mu+\left(\pounds_{\xi_{(2)}} +
\pounds^2_{\xi_{(1)}}\right)u^\mu_{(0)}
+2\pounds_{\xi_{(1)}}\delta u^\mu\;.
\end{equation}
Using (\ref{eq:4v}) and (\ref{eq:liedev2}) this gives:
\begin{eqnarray}\label{eq:v02}
\fl \tilde{v}^0_{(2)} & = &
v^0_{(2)}-\frac{a'}{a}\alpha_{(2)}-\alpha_{(2)}^\prime
+\alpha_{(1)}\left[2\left(v^{0\prime}_{(1)}-
\frac{a'}{a}v^0_{(1)}\right)+
\left(2\frac{a^{\prime
2}}{a^2}-\frac{a^{\prime\prime}}{a}\right)\alpha_{(1)}
\right.\nonumber\\
\fl &&\left.+\frac{a'}{a}\alpha^\prime_{(1)}-
\alpha^{\prime\prime}_{(1)}\right] +\xi^i_{(1)}\left(2 v^0_{(1),i}
-\frac{a'}{a}\alpha^{(1)}_{,i}-\alpha^{(1)\prime}_{,i}\right) \\
\fl &&+\alpha^\prime_{(1)}\left(\alpha^\prime_{(1)}-2v^0_{(1)}\right) 
-2\alpha^{(1)}_{,i}v^i_{(1)}
+\alpha^{(1)}_{,i}\xi^{i\prime}_{(1)}\;;\nonumber\\ 
\label{eq:vi2}
\fl \tilde{v}^i_{(2)} & = & v^i_{(2)} 
 -\beta_{(2)}^{\prime ,i} - d_{(2)}^{i\prime}
+\alpha_{(1)}\left[
2\left(v^{i\prime}_{(1)}-\frac{a'}{a}v^i_{(1)}\right)
-\left(\xi^{i\prime\prime}_{(1)}
-2\frac{a'}{a}\xi^{i\prime}_{(1)}\right)\right]\nonumber
\\  \fl & & 
+\xi^j_{(1)}\left(2v^i_{(1),j}-\xi^{i\prime}_{(1),j}\right)
-\xi^i_{(1),j} \left( 2 v^j_{(1)} -\xi^{j\prime}_{(1)}\right)
+\xi^{i\prime}_{(1)} 
\left( 2 \psi_{(1)} +\alpha_{(1)}^\prime\right)\;;
\end{eqnarray}
for the time and the space  components respectively.
Again, the 4-velocity $u^\mu$ is subject to $u^\mu u_\mu=-1$, which
gives (\ref{eq:v0psi2}); therefore (\ref{eq:v02}) reduces to 
(\ref{eq:psi2}).

%%%%%%%%%%%%%%%%%%%%%%%%%%%%%%%%%%%%%%%%%%%%%%%%%%%%%%%%%%%%%%%%%%%%%%

\subsubsection{Transforming from the
 synchronous to the Poisson gauge}

For this example, let us consider the simplified case in which only
scalar first order perturbations are present as initial conditions for
the second order problem. In the first order analysis presented above,
this corresponds to having $\chi_{{\scriptscriptstyle \rm S}\
ij}^{(1)\bot}= \chi_{{\scriptscriptstyle \rm S}\
ij}^{(1)\top}=v^i_{(1)\perp}=0$, and thus
$d_i^{(1)}=\omega^{(1)}_{{\scriptscriptstyle \rm P}\ i}
=\chi_{{\scriptscriptstyle \rm P}\ ij}^{(1)\bot}=0$. The second order
vector and tensor perturbations are however non-vanishing as the
dynamical coupling of the modes makes them grow when non-linear terms
are considered in the evolution equation.  We consider this
restriction just for the sake of simplicity and because it describes a
physically interesting situation. The more general transformation
expressions follow straightforwardly from 
(\ref{eq:psi2})--(\ref{eq:chi2}), (\ref{eq:mut2}), and (\ref{eq:vi2}).

Transforming from the synchronous to the Poisson gauge, the expression
for $\psi^{(2)}_{\scriptscriptstyle \rm P}$ can be easily obtained
from (\ref{eq:psi2}), using (\ref{iii}) and the condition
$d_i^{(1)}=0$ to express all the first order quantities in terms of
$\beta^{(1)}$:
\begin{eqnarray}
\label{eq:psiP}
\psi^{(2)}_{\scriptscriptstyle \rm P}=&\beta_{(1)}^{\prime}\left[
 \beta_{(1)}^{\prime\prime\prime}
+5\frac{a'}{a}\beta_{(1)}^{\prime\prime}
+\left(\frac{a''}{a} +\frac{a^{\prime 2}}{a^2}\right) 
\beta_{(1)}^{\prime}\right]\nonumber\\
&+\beta_{(1)}^{ ,i}\left(
\beta^{(1)\prime\prime}_{,i}
+\frac{a'}{a}\beta^{(1)\prime}_{,i}\right)
+2\beta_{(1)}^{\prime\prime 2} 
+\alpha^{(2)\prime}+\frac{a'}{a}\alpha^{(2)}\;.
\end{eqnarray}

For $\omega^{(2)}_{{\scriptscriptstyle \rm P}\ i}$ and
$\phi^{(2)}_{\scriptscriptstyle \rm P}$ we get:
\begin{equation}\label{eq:omegaP}
\fl \omega^{(2)}_{{\scriptscriptstyle \rm P}\ i}=
-2\left(
2\phi^{(1)}_{\scriptscriptstyle \rm S} 
+\beta_{(1)}^{\prime\prime} 
-\frac{2}{3}\nabla^2\beta_{(1)}\right)\beta^{(1)\prime}_{,i}
-2\beta^{(1)\prime}_{,j}\beta^{(1),j}_{,i}-
\alpha^{(2)}_{,i} +\beta^{(2)\prime}_{,i} +d^{(2) \prime}_i\;;
\end{equation}
\begin{eqnarray}\label{eq:phiP}
\fl \phi^{(2)}_{\scriptscriptstyle \rm P}& =
 & \phi^{(2)}_{\scriptscriptstyle
\rm S}
+\beta^{\prime}_{(1)}\left[
2\left(\phi_{\scriptscriptstyle
\rm S}^{(1)\prime}+
2\frac{a'}{a}\phi^{(1)}_{\scriptscriptstyle
\rm S}\right)
-\left(\frac{a''}{a}
+\frac{a^{\prime 2}}{a^2}\right)\beta^{\prime}_{(1)}
-\frac{a'}{a} \beta_{(1)}^{\prime\prime}\right]
\nonumber \\
\fl & &
-\frac{1}{3}\left(-4\phi^{(1)}_{\scriptscriptstyle
\rm S}+\beta_{(1)}^{\prime}\partial_0
+\beta^{,i}_{(1)}\partial_i
+4\frac{a'}{a}\beta^{\prime}_{(1)}
+\frac{4}{3}\nabla^2\beta_{(1)}\right)\nabla^2\beta_{(1)}
\\ 
\fl && \nonumber
+\beta^{,i}_{(1)}\left( 2\phi^{(1)}_{{\scriptscriptstyle
\rm S} ,i}-\frac{a'}{a}\beta^{(1)\prime}_{,i}\right)
+\frac{2}{3}\beta^{(1)}_{,ij}\beta^{,ij}_{(1)}
-\frac{a'}{a}\alpha_{(2)} -\frac{1}{3}\nabla^2\beta_{(2)}\;.
\end{eqnarray}

For $\chi^{(2)}_{{\scriptscriptstyle \rm P}\ ij}$ we obtain:
\begin{eqnarray}
\chi^{(2)}_{{\scriptscriptstyle \rm P}\ ij} & = & 
\chi^{(2)}_{{\scriptscriptstyle \rm S}\ ij}
+2\left(\frac{4}{3}\nabla^2\beta_{(1)}
-4\phi^{(1)}_{\scriptscriptstyle
\rm S} -
\beta^{\prime}_{(1)}\partial_0-\beta_{(1)}^{,k}\partial_k\right){\rm
D}_{ij} \beta_{(1)}
 \nonumber \\
& & \label{eq:chiP}
-4\left(\beta^{(1)}_{,ik}\beta^{,k}_{(1),j}
-\frac{1}{3}\delta_{ij}\beta^{(1)}_{,lk}\beta_{(1)}^{,lk}\right)
+2\left(d^{(2)}_{(i,j)} +{\rm D}_{ij}\beta^{(2)}\right)\; .
\end{eqnarray} 

Given the metric perturbations in the synchronous gauge, these
constitute a set of coupled equations for the second order parameters
of the transformation, $\alpha^{(2)},\beta^{(2)}$, and $d_i^{(2)}$,
and the second order metric perturbations in the Poisson gauge,
$\psi_{\scriptscriptstyle \rm P}^{(2)}$, $\omega_{{\scriptscriptstyle
\rm P}\ i}^{(2)}$, $\phi_{\scriptscriptstyle \rm P}^{(2)}$, and
$\chi^{(2)}_{{\scriptscriptstyle \rm P}\ ij}$.  This system can be
solved in the following way.  Since in the Poisson gauge
$\partial^i\chi^{(2)}_{{\scriptscriptstyle \rm P}\ ij}=0$, we can use
the fact that $\partial^i \partial^j \chi^{(2)}_{{\scriptscriptstyle
\rm P}\ ij}=0$ and the property $\partial^id_i^{(1)}=0$, together with
(\ref{eq:chiP}), to obtain an expression for
$\nabla^2\nabla^2\beta^{(2)}$, from which $\beta^{(2)}$ can be
computed:
\begin{eqnarray}\label{beta2}
\fl \nabla^2\nabla^2\beta_{(2)} & = & -\frac{3}{4}
\chi^{(2),ij}_{{\scriptscriptstyle \rm S}\ ij}+
6\phi_{\scriptscriptstyle \rm S}^{(1),ij}
\beta^{(1)}_{,ij} -2 \nabla^2\phi_{\scriptscriptstyle \rm S}^{(1)} 
\nabla^2\beta_{(1)}
+8  \phi_{\scriptscriptstyle \rm S}^{(1),i} \nabla^2\beta^{(1)}_{,i}
\nonumber\\
\fl &&
+4 \phi_{\scriptscriptstyle \rm S}^{(1)} \nabla^2\nabla^2\beta_{(1)} 
+4\nabla^2\beta^{(1)}_{,ij}\beta_{(1)}^{,ij}
-\frac{1}{6}\nabla^2\beta_{(1)}^{,i} \nabla^2\beta^{(1)}_{,i}
+\frac{5}{2}\beta_{(1)}^{,ijk} \beta^{(1)}_{,ijk}\nonumber\\
\fl &&-\frac{2}{3}\nabla^2\beta_{(1)} \nabla^2\nabla^2\beta_{(1)}
+\frac{3}{2}\beta_{(1)}^{,ij\prime} \beta^{(1)\prime}_{,ij} 
-\frac{1}{2}\nabla^2\beta_{(1)}^{\prime} \nabla^2\beta_{(1)}^{\prime}\\
\fl && +2\beta_{(1)}^{,i\prime} \nabla^2\beta^{(1)\prime}_{,i}
+\beta_{(1)}^{\prime} \nabla^2\nabla^2\beta_{(1)}^{\prime} 
+\beta_{(1)}^{,i} \nabla^2\nabla^2\beta^{(1)}_{,i}\;.\nonumber
\end{eqnarray}
Then, using the condition $\partial^i\chi^{(2)}_{{\scriptscriptstyle
\rm P}\ ij}=0$ and substituting $\beta^{(2)}$ 
we obtain an equation for  $d^{(2)}_i$: 
\begin{eqnarray}
\nabla^2 d^{(2)}_i & = &-\frac{4}{3}\nabla^2\beta^{(2)}_{,i}
-\chi^{(2),j}_{{\scriptscriptstyle \rm S}\ ij}+
8\phi_{\scriptscriptstyle \rm S}^{(1),j}{\rm D}_{ij}\beta_{(1)} 
+\frac{16}{3}\phi_{\scriptscriptstyle \rm S}^{(1)}
\nabla^2\beta^{(1)}_{,i}\nonumber\\
&&+\frac{2}{3}\nabla^2\beta_{(1)}^{,j}\beta^{(1)}_{,ij}
+\frac{10}{3}\beta_{(1)}^{,jk}\beta^{(1)}_{,ijk}
-\frac{8}{9}\nabla^2\beta_{(1)} \nabla^2\beta^{(1)}_{,i}\\
&& +2\beta_{(1)}^{,j\prime}{\rm D}_{ij}\beta_{(1)}^{\prime} 
+\frac{4}{3}\beta_{(1)}^{\prime}\nabla^2\beta^{(1)\prime}_{,i}
+\frac{4}{3}\beta_{(1)}^{,j}\nabla^2\beta^{(1)}_{,ij}.\nonumber
\end{eqnarray}
Finally, using $\partial^i\omega_{{\scriptscriptstyle \rm P}\
i}^{(2)}=0$ and substituting $\beta^{(2)}$, we get an equation for
$\alpha^{(2)}$:
\begin{eqnarray}
\nabla^2\alpha_{(2)} & = & \nabla^2\beta_{(2)}^{\prime}
-2\left(2\phi^{(1),i}_{\scriptscriptstyle \rm S}
+\beta_{(1)}^{\prime\prime,i} +\frac{1}{3}\nabla^2\beta_{(1)}^{,i}
\right)\beta^{(1)\prime}_{,i}
\nonumber \\
& &
-2\left(2\phi^{(1)}_{\scriptscriptstyle \rm S}
+\beta_{(1)}^{\prime\prime} -\frac{2}{3}\nabla^2\beta_{(1)}
\right)\nabla^2\beta_{(1)}^{\prime}
-2\beta_{(1)}^{,ij}\beta^{(1)\prime}_{,ij}\;.
\end{eqnarray}
Having obtained, at least implicitly, all the parameters of the gauge
transformation to second order, one can in principle compute the metric
perturbations in the Poisson gauge from (\ref{eq:psiP})--(\ref{eq:chiP}).

Similarly, once the parameters are known, 
the perturbations in any scalar and 4-vector, and in
particular those in the 
energy density and in the 4-velocity of matter, follow trivially from
(\ref{eq:mut2})--(\ref{eq:vi2}).

%%%%%%%%%%%%%%%%%%%%%%%%%%%%%%%%%%%%%%%%%%%%%%%%%%%%%%%%%%%%%%%%%%%%%%
%%%%%%%%%%%%%%%%%%%%%%%%%%%%%%%%%%%%%%%%%%%%%%%%%%%%%%%%%%%%%%%%%%%%%%
%%%%%%%%%%%%%%%%%%%%%%%%%%%%%%%%%%%%%%%%%%%%%%%%%%%%%%%%%%%%%%%%%%%%%%

\section{Conclusions}
\setcounter{equation}{0}

In this paper we have studied the problem of gauge dependence in
relativistic perturbation theory, considering perturbations of
arbitrary order in a geometrical perspective.  In fact, the problem
itself is of a purely geometrical nature, dealing with the
arbitrariness in the mapping between the physical spacetime and the
background unperturbed one.  Since no dynamics is involved, the
formalism developed here can actually find application not only in
general relativity, but in any spacetime theory.  In considering a
specific example, we have assumed a flat Robertson--Walker background,
and derived the second order transformation between the well-known
synchronous gauge \cite{bi:li46}, and the Poisson (generalized
longitudinal) gauge discussed in  \cite{bi:bertschinger}.

In linearized perturbation theory a gauge transformation is generated
by an arbitrary vector field $\xi_{(1)}$, defined on the background
spacetime, and associated with a one-parameter group of
diffeomorphisms (a flow): the gauge transformation of the perturbation
$\delta T$ of a tensor field $T$ is then given by the Lie derivative
$\pounds_{\xi_{(1)}} T_0$ of the background field $T_0$.  However, in
considering a gauge transformation  from an exact point of view, 
we have found that it is not represented by a flow, but rather by a
more general one-parameter family of diffeomorphisms. The question
then was, how can we approximate the latter to a given \hbox{order
$n$?} To this end, we have developed in section 2 the necessary
mathematical formalism.  First, we have introduced certain families of
mappings, dubbed knight diffeomorphisms of rank $n$, defined by 
(\ref{eq:knightn}).  Then, in Theorem 2, we have proved that any
one-parameter family of diffeomorphisms may always be approximated, to
order $n$, by a knight diffeomorphism of rank $n$.  This result
(which confirms a claim in \cite{bi:GL}) is
fundamental for gauge transformations of order $n$, as it guarantees
that they are correctly represented by knight diffeomorphisms of the
same rank.  From the applicative point of view, Lemma 2 is thus all we
need to have a generating formula for the gauge transformation to an
arbitrary order, (\ref{lemma2applied}).  Since a knight
diffeomorphism of rank $n$ is basically the composition of $n$ flows,
and is thus generated by $n$ vector fields $\xi_{(1)},\ldots,
\xi_{(n)}$, a gauge transformation of order $n$ for the $n$-th order
perturbation $\delta^n T$ of a tensor field $T$ involves an
appropriate combination of the Lie derivatives along
$\xi_{(1)},\ldots, \xi_{(n)}$ of $T_0,\ldots,\delta^{n-1}T$.

Gauge transformations found their main application in considering
the time evolution of
perturbations of a given background spacetime: in a subsequent
paper\cite{bi:MMB} 
we shall look at second order perturbations of an Einstein de Sitter
universe, comparing results in the synchronous and the Poisson gauges,
thus applying the results presented  in section 5. 
Beyond these applications, there are many topics that 
 we have not touched upon here which, in a way or another,
are related to gauge transformations, and become even more cumbersome
in the non-linear case. We mention only a couple of them.  We have
implicitly assumed the applicability of the perturbative method; also,
we have not considered the problem of eliminating spurious gauge
modes.
In particular, in our cosmological example, the synchronous
gauge as defined in section 5 should actually be regarded as a {\it
class\/} of point identification maps \cite{bi:KS}.

A final issue we would like to mention is that of gauge-invariant
quantities. In relation to this, we have first defined gauge
invariance in an exact sense, and then given the conditions for the
gauge invariance of a tensor field $T$ to an arbitrary perturbative
order $n$. However, we have not faced the problem of finding or
constructing such quantities. In particular, in considering
cosmological perturbations, it would be useful to have at hand a set
of second order gauge-invariant variables defined {\it \`a la\/}
Bardeen \cite{bi:JB}. However, given the gauge-dependent metric
perturbations and their transformation rules presented in section 5, the
construction of such variables seems impractical. Moreover, it is far
from obvious that a complete set giving a full second order
description exists at all, as is the case at first
order \cite{bi:JB,bi:goode}.  Another possibility is to look for
covariant quantities \cite{bi:EB,bi:BDE}: in order to be second order
gauge-invariant these should vanish in the background and at first
order.  Assuming purely scalar first order perturbations, two examples
are the magnetic part of the Weyl tensor and the vorticity of the
4-velocity of matter.  Other second order gauge-invariant quantities
can be defined by taking products of first order gauge-invariant
tensors that vanish in the background.  In particular, this is the
case for scalars such as $E_{\mu\nu}E^{\mu\nu}$, where $E$ is the
electric part of the Weyl tensor.  Once again, it seems difficult that
a complete set of such variables could even exist.  Nevertheless, it
is worth pointing out that quantities which are quadratic in first
order gauge-invariant variables are useful in specific problems; for
example they may intervene in the construction of effective energy
momentum tensors of perturbations, which are important for the study
of back reaction problems \cite{taub,MAB}. Another 
 possible application of
the formalism developed here can be the study of the gauge dependence
of these quantities.

%%%%%%%%%%%%%%%%%%%%%%%%%%%%%%%%%%%%%%%%%%%%%%%%%%%%%%%%%%%%%%%%%%%%%%
%%%%%%%%%%%%%%%%%%%%%%%%%%%%%%%%%%%%%%%%%%%%%%%%%%%%%%%%%%%%%%%%%%%%%%
%%%%%%%%%%%%%%%%%%%%%%%%%%%%%%%%%%%%%%%%%%%%%%%%%%%%%%%%%%%%%%%%%%%%%%

\ack

This work has been partially supported by the Italian MURST; MB acknowledges
ICTP and INFN for financial support.  MB 
thanks John Friedman for a stimulating conversation.  
SS is grateful to Dennis W.\ Sciama for hospitality at the
Astrophysics Sector of SISSA.
We  thank Eanna Flanagan and Bernard F. Schutz who, 
after this paper was circulated in electronic form (gr-qc/9609040),
drawn our attention to references \cite{bi:GL,bi:FW}  and  \cite{bi:schutz} 
(see also references therein).

%%%%%%%%%%%%%%%%%%%%%%%%%%%%%%%%%%%%%%%%%%%%%%%%%%%%%%%%%%%%%%%%%%%%%%
%%%%%%%%%%%%%%%%%%%%%%%%%%%%%%%%%%%%%%%%%%%%%%%%%%%%%%%%%%%%%%%%%%%%%%
%%%%%%%%%%%%%%%%%%%%%%%%%%%%%%%%%%%%%%%%%%%%%%%%%%%%%%%%%%%%%%%%%%%%%%

%%%%%%%%%%%%%%%%%%%%%%%%%%%%%%%%%%%%%%%%%%%%%%%%%%%%%%%%%%%%%%%%%%%%%%
%%%%%%%%%%%%%%%%%%%%%%%%%%%%%%%%%%%%%%%%%%%%%%%%%%%%%%%%%%%%%%%%%%%%%%
%%%%%%%%%%%%%%%%%%%%%%%%%%%%%%%%%%%%%%%%%%%%%%%%%%%%%%%%%%%%%%%%%%%%%%

\end{document}